\documentclass[preprint,12pt]{elsarticle}

\usepackage{graphicx}

\usepackage{amssymb,amsmath}

\journal{Physica A: Statistical Mechanics and its Applications}

\begin{document}

\begin{frontmatter}



\title{Change of Scaling and Appearance of Scale-Free Size Distribution in Aggregation Kinetics by Additive Rules}


\author{Yuri G. Gordienko}

\address{G.V.Kurdyumov Institute for Metal Physics, National Academy of Sciences of Ukraine, 36 Academician Vernadsky Blvd, Kiev 03680, Ukraine}

\begin{abstract}
The idealized
general model of aggregate growth is considered on the basis of the simple
additive rules that correspond to one-step aggregation process.
The two idealized cases were analytically
investigated and simulated by Monte Carlo method
in the Desktop Grid distributed computing environment
to analyze ``pile-up'' and ``wall'' cluster distributions in different aggregation scenarios.
Several aspects of aggregation kinetics (change of scaling, change of size distribution type,
and appearance of scale-free size distribution) driven by ``zero cluster size'' boundary condition were
determined by analysis of evolving cumulative distribution functions.
The ``pile-up'' case with a
\textit{minimum} active surface (singularity) could imitate piling up aggregations of
dislocations, and the case with a \textit{maximum} active surface could imitate
arrangements of dislocations in walls. The change of scaling law (for pile-ups and
walls) and availability of scale-free distributions (for walls) were
analytically shown and confirmed by scaling, fitting, moment, and
bootstrapping analyses of simulated probability density and cumulative distribution functions.
The initial ``singular'' \textit{symmetric} distribution of pile-ups evolves by the ``infinite'' diffusive scaling law
and later it is replaced by the other ``semi-infinite'' diffusive scaling
law with \textit{asymmetric} distribution of pile-ups. In contrast, the initial ``singular''
\textit{symmetric} distributions of walls initially evolve by the diffusive scaling law
and later it is replaced by the other ballistic (linear) scaling law with
\textit{scale-free} exponential distributions without distinctive peaks.
The conclusion was made as to possible applications of such approach for scaling, fitting, moment, and bootstrapping analyses
of distributions in simulated and experimental data.
\end{abstract}


\end{frontmatter}


\section{Introduction}
\label{}
Many aggregation phenomena in natural
processes take place by exchange of solitary agents (monomers) between their
aggregates (clusters): phase ordering \cite{LifshitzSlyozovRU},
atom deposition \cite{Zangwill}, stellar
evolution \cite{Chandrasekhar},
growth and distribution of assets \cite{Ispolatov}, and city population
even \cite{Leyvraz}. 
In materials science the hierarchical defect substructures that were observed experimentally in
deformed metals and alloys appear as a result of some aggregation processes
among solitary crystal defects.
As a result, the hierarchical defect substructures 
can demonstrate the self-affine geometry on many scales. 
In fact, the fractal analysis of fractured surfaces by projective covering and box-counting method shows that the
fractured surface can be depicted not only by one fractal dimension, but
also by multifractal spectrum \cite{Mandelbrot,Stach2003theory,Stach2003exp}. At the same time, surface roughness
profiles of periodically deformed Al \cite{GordienkoSPIE,GordienkoPhysMez2007,GordienkoMSF2008},
slip line morphology in Cu \cite{Kleiser,Neuhauser},
and dislocation patterns in Cu after tensile \cite{Zaiser}
also demonstrate the self-similar features on many scales.
Recently, transition from homogeneous dislocation arrangement to
scale-invariant structure was described by the statistical model of
noise-induced transition  \cite{Hahner}. Several other models and
theories were proposed to explain the scale-invariant behavior
of crystal defect aggregations \cite{GordienkoPhilMag1994}
that possibly lead to self-affine geometry of fractured surfaces \cite{Ananthakrishna}.

The general model of aggregate growth on the basis of the simple
additive rules that correspond to one-step aggregation process
and its scaling properties are of great interest in this physical context.
In such one-step aggregation processes monomers can leave one cluster and
attach to another. Usually these exchange processes are described by an
exchange rate kernel $K(i,\mbox{ }j)$, i.e. by the rate of transfer of monomers
from a cluster of size $i$ (detaching event) to a cluster of size $j$ (attaching
event). Generally, the rate of monomer exchange between two clusters depends
on their active interface surfaces that are dependent on their sizes,
morphology (line, plane, disk, sphere, fractal, etc), probability of
detaching and attaching events, etc.

Sometimes there is the preferable direction for exchanges, i.e. with
asymmetric exchange kernels, $K(i,\mbox{ }j)\ne \mbox{ }K(\mbox{ }j,i)$,
like in coalescence processes in Lifshitz-Slyozov-Wagner theory \cite{LifshitzSlyozovRU}, where
big clusters ``eat'' smaller ones. The exchange rate kernel $K(i,\mbox{ }j)$
is defined by the product of the rate at which monomer detach from a cluster
of size $i$ and the rate at which this monomer reach another cluster of size
$j$.

In Leyvraz-Redner scaling theory of aggregate growth \cite{Leyvraz} cities $A_i $ of
size $i$ evolve according to the following rule:

\begin{equation}
\label{eq1}
A_i +A_j \buildrel {K\left( {i;j} \right)} \over \longrightarrow A_{i-1}
+A_{j+1} ,
\end{equation}

where $K(i,\mbox{ }j)$ is the exchange rate. That is, monomer (one person)
leaves some of cities $A_i $ of population $i$ and arrive to some of cities
$A_j $ of population $j$. This can be considered as the generalized rule for the
theory of growth and distribution of assets \cite{Ispolatov}, if one can assume that
$A_i $ are persons with asset volume of $i$.

Below the idealized general model of aggregate growth is proposed on the
basis of this approach. The main aim of the work is to use the most
profound features of aggregation kinetics
and to find the simplest factors that can cause
the observed self-affine properties of the aggregating system of solitary agents
(monomers) and their aggregates (clusters). In this context, the numerous
complex details of the real crystal defect aggregation processes will be
hidden behind the idealized and simplified conditions only to emphasize the
most general precursors of scale-invariant behavior of such complex systems.

\section{Model}
\label{Model}
Here detaching and attaching processes are considered \textit{separately} that in the
general case could be characterized by different rates.
    The proposed model significantly differs by this aspect from
    the other well-known aggregation models in Leyvraz-Redner scaling theory of aggregate growth \cite{Leyvraz},
    Ben-Naim-Krapivsky theory for exchange driven growth \cite{ben2003exchange},
    Lin-Ke theory for migration-driven aggregation \cite{ke2002kinetics,lin2003kinetics,lin2005exchange},
    where detaching and attaching processes are considered \textit{together}
    in the formalism of the linked Smolukhovski nonlinear equations \cite{Smoluchowski1916}.
Consequently, the different detach product kernel $K_d (n)=k_d S\left( n \right)$ and attach
product kernel $K_a (n)=k_a S_a \left( n \right)$ are taken into account,
where $k_d $ and $k_a $ are the measures of activation of detaching and
attaching processes, $n$ is the number of monomers in a cluster. In natural
processes $k_d $ is usually determined by energy barrier for detachment from
cluster and $k_a $~--- by probability for attachment of migrating monomer to
another cluster which in turn determined by kind of migration (instant hops
from cluster to cluster, ballistic motion, random walking, or their
combinations). $S_d \left( n \right)=s_d n^\alpha $ and $S_a \left( n
\right)=s_a n^\beta $ are the active surfaces of clusters, where $\alpha $
and $\beta $ --- exponents depending on the morphology of cluster (for
example $\alpha =1$ for linear clusters and $\alpha =2 \mathord{\left/
{\vphantom {2 3}} \right. \kern-\nulldelimiterspace} 3$ for spherical
clusters, and $\alpha =\beta $ in the simplest case of clusters with the
same morphology), $s_d $ and $s_a $~--- the constants depending on the
morphology of cluster and geometry of neighborhood (for example $s_d =1$ for
linear aggregates and $s_d =\sqrt[3]{36\pi }$ for spherical aggregates, and
$s_d =s_a =s$ in the simplest case of clusters with the same morphology and
neighborhood). The portion of clusters $f(n,t)$ with $n$ monomers at time $t$
evolves according to the following equation:

\begin{equation}
\label{eq2}
\begin{split}
\frac{\partial f(n,t)}{\partial t}&= K_d (n+1)f(n+1,t)+K_a (n-1)f(n-1,t)-\\
                                    &-K_d (n)f(n,t)-K_a (n)f(n,t)
\end{split}
\end{equation}

Actually this is the well-known master equation for so-called ``one-step
process'', i.e. a continuous time stochastic Markov process, which range
consists of integer $n$, and where transitions allowed only between adjacent
integers \cite{vanKampen}.

The general model is based on the assumption that the migration time
(movement of a monomer from one cluster to other) is much lower than the
detachment (or attachment) time of a monomer from (to) a cluster. That is
why there are no free monomers, and range of $n$ is half-infinite
($n=2,3,{\ldots}$). At the same time the total number of monomers is assumed to be constant
and no generation sources exist.
It should be taken into account that the more consistent
system of evolution equations and more realistic coupling between migration
and detachment (attachment) times should be used for rigorous comparison of
this general model and each case of the aforementioned natural processes.



In an asymptotic regime of high values of $n$ equation (\ref{eq2}) goes to:

\begin{equation}
\label{eq3}
\frac{\partial f\left( {n,t} \right)}{\partial t}\approx \frac{\partial
\left( {D_1 \left( n \right)f\left( {n,t} \right)} \right)}{\partial
n}+\frac{\partial ^2\left( {D_2 \left( n \right)f\left( {n,t} \right)}
\right)}{\partial n^2},
\end{equation}

which is the one-variable Fokker-Planck equation (FPE)  \cite{Fokker,Planck} in general form
with time-independent drift
    $D_1 (n)=K_d (n)-K_a (n)= s\left( {n^\alpha k_d - n^\beta k_a }\right)$
and diffusion
    $D_2 (n)={1 \over 2}(K_d (n) + K_a (n))={1 \over 2}(n^\alpha sk_d + n^\beta sk_a)$
coefficients in the proposed model in a general case, when $K_d(n) \neq K_a(n)$ ($\alpha \neq \beta$,  $k_d \neq k_a$).

The main difference between this formulation and
  the other aforementioned models \cite{Leyvraz,ben2003exchange,ke2002kinetics,lin2003kinetics,lin2005exchange}
is the more general scenario with detaching and attaching as separate
events is taken into account here.
%
%
    In contrast to them the proposed model allows us:
    \begin{enumerate}
      \item to express explicitly the availability of drift and diffusion components of migration-driven aggregation through the terms of the same name in FPE (\ref{eq3}),
      \item to emphasize conditions of symmetric/asymmetric migration-driven aggregation, namely, for asymmetric (biased) one with $D_1(n)=K_d(n)-K_a(n)= s\left( {n^\alpha k_d - n^\beta k_a }\right) \neq 0$,
          i.e. when $\alpha \neq \beta$ and $k_d \neq k_a$,
          for symmetric (unbiased) one with $D_1(n)=0$, i.e. when $\alpha = \beta$ and $k_d = k_a$;
      \item to find the exact asymptotic non-stationary solutions for unbiased migration-driven aggregation \cite{gordienkoIJMPB2012};
      \item to analyze the common symmetry properties of FPE (\ref{eq3}) and propose the ways to find the exact non-stationary solutions for the more general cases \cite{GatsenkoCGW10,gordienkoECF2008}, for example, for biased migration-driven aggregation.
    \end{enumerate}

\begin{figure}[htbp]
\centerline{
\includegraphics[width=1.6in,height=.5in]{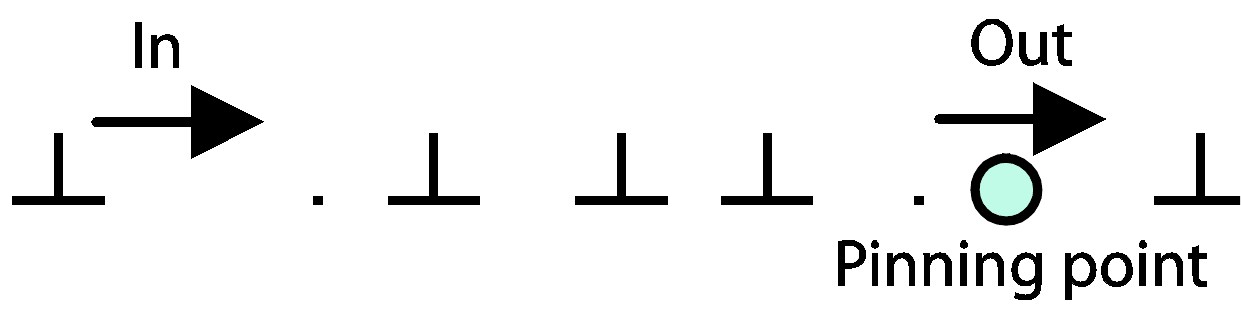}
\hspace{2cm}
\includegraphics[width=.9in,height=1.2in]{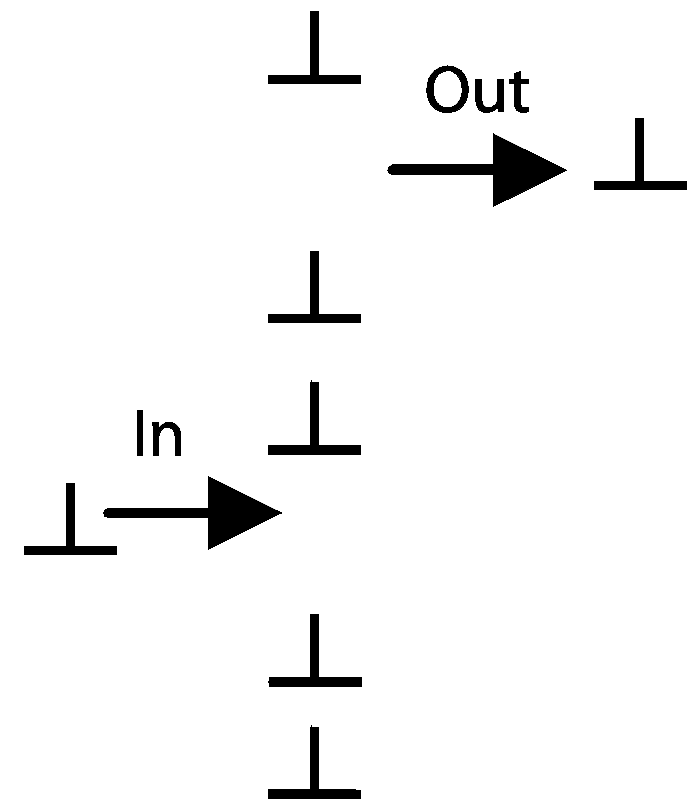}
}
\centerline{a) \hspace{5cm} b)}
\caption{Examples of (a) minimum active surface --- pile-up of dislocations:
it is possible to go in/out pile-up only through left/right ends of the
pile-up; and (b) maximum active surface --- wall of dislocations: it is
possible to go in/out wall in any place.}
\label{Examples}
\end{figure}

The case $\alpha =0$ corresponds to the clusters with the minimum active
surface (``singularity''' of the active surface), namely for clusters with
constant numbers of active monomers independent of the whole number of
monomers $n$ in it. For example, the ends of a line cluster can be its 2 active
points for detaching and attaching.
    %
    %
      Such configurations take place in dequeues, queues
      (2 active points), stacks (1 active point), etc. that are well-known in computer science,
      especially in queuing and waiting theories \cite{knuth,bhat2008}.
In the context of plastic
deformation of metals, ``pile-up'' aggregations of dislocations of a regular
crystalline structure can be depicted by this scenario (see Fig.~\ref{Examples}a).
      The similar case was considered also in Ke-Lin theory
      for migration-driven aggregation \cite{ke2002kinetics},
      but on the basis of the linked Smolukhovski nonlinear equations \cite{Smoluchowski1916}.

The case $\alpha =1$ corresponds to the clusters with the maximum active
surface, namely for clusters where any monomer could be detached and free
monomer could attached to any place. Again, in plastically
deformed metals, ``wall'' aggregations of dislocations of
a regular crystalline structure can be depicted
by this scenario (see Fig.~\ref{Examples}b). In Leyvraz-Redner scaling theory of
aggregate growth of city population \cite{Leyvraz} it can correspond to linear
dependence of arrival or departure rates as a function of city population.
Again, the similar case was considered also in Lin-Ke theory
for migration-driven aggregation \cite{lin2003kinetics},
but on the basis of the linked Smolukhovski nonlinear equations \cite{Smoluchowski1916}.

The case $0<\alpha <1$ corresponds to the clusters with the bulk volume of
monomers shielded by active surface. For example, circumference of disk,
perimeter of fractal, outer layer of sphere can be the correspondent active
surfaces for detaching and attaching. In solid state physics, such
configurations take place in regular arrangements (compact like voids and
spare like fractals) of point-like defects of a crystalline structure.
This case was investigated in the other work \cite{gordienkoIJMPB2012},
where the exact non-stationary solutions were found under conditions of unbiased migration-driven aggregation with the conserved number of monomers.

Below the symmetric case $K_d(n) = K_a(n)$ is considered,
which corresponds to unbiased migration-driven aggregation
with the equiprobable activation of detaching
and attaching processes. For example, for periodic tensile deformation it
can correspond to the equiprobable activation of detaching and attaching
dislocations from opposite sides of wall. Finally, it means the absence of
the drift term in (\ref{eq3}).

\paragraph{Pile-up - minimum active surface}
For the case $\alpha =0$ clusters have the minimum active
surface and under condition $K_d(n) = K_a(n)$ one can get the well-known heat equation, where $a=sk_a$,

\begin{equation}
\label{eq4_heat}
\frac{\partial f_p\left( {n,t} \right)}{\partial t}=a\frac{\partial ^2f_p\left(
{n,t} \right)}{\partial n^2},
\end{equation}

which has numerous particular solutions that are dependent on the initial
and boundary conditions. In the reality $k_d \ne k_a $, and the aggregation
process is described by the homogeneous heat equation with space-independent
drift and diffusion coefficients. It is well-known fact that it leads to ``a
diffusive-like kinetic universality class''.

Below for illustrative purpose we consider the time evolution of the initial
singular distribution of clusters of the same size ($n_0 )$ that exchange
the monomers. It is actually the first boundary value problem for domain
$0<n<\infty $, $f\left( {0,t} \right)=0$, $f\left( {n,0} \right)=\delta
\left( {n-n_0 } \right)$, where $\delta \left( n \right)$ is a Dirac delta
function. This problem has the well-known solution:

\begin{equation}
\label{eq5}
f_p\left( {n,t} \right)=\frac{1}{2\sqrt {\pi at} }\left\{ {\exp \left[
{-\frac{\left( {n-n_0 } \right)^2}{4at}} \right]-\exp \left[ {-\frac{\left(
{n+n_0 } \right)^2}{4at}} \right]} \right\}.
\end{equation}

In practice, $f\left( {n,t} \right)$ could be experimentally determined for
high values of $n$ and $t$. That is why $f\left( {n,t} \right)$ is of interest for
large values of $n \gg n_0 $ and $t \gg t_{0,p} = n n_0 / a$:

\begin{equation}
\label{eq6}
f_{p,t>t_{0,p}}(n,t) \to \frac{nn_0 }{2\sqrt \pi \left( {at} \right)^{3
\mathord{\left/ {\vphantom {3 2}} \right. \kern-\nulldelimiterspace} 2}}\exp
\left[ {-\frac{n^2}{4at}} \right].
\end{equation}

The essential point is that on the initial stage of this one-step
aggregation process the probability distribution function $f\left( {n,t}
\right)$ does not ``feel'' the boundary condition $f\left( {0,t} \right)=0$.
From the physical point of view the cluster size cannot reach
the 0-boundary (zero cluster size) and equation cannot ``feel'' the boundary condition immediately,
but only after some time $t_{0,p}$, when the first
cluster disappear due to detaching monomers. It means that on the initial
stage $t<t_{0,p}$ the aforementioned problem formulated as the classic Cauchy
problem for \textit{infinite} domain $-\infty <n<\infty $ with the same initial condition
$f\left( {n,0} \right)=\delta \left( {n-n_0 } \right)$ and the well-known
solution:

\begin{equation}
\label{eq7}
f_{p,t<t_{0,p}}(n,t) = \frac{1}{2\sqrt {\pi at} }\exp \left[
{-\frac{\left( {n-n_0 } \right)^2}{4at}} \right].
\end{equation}

Thus, in the context of the same process
we have change of size distribution (from (\ref{eq7}) to (\ref{eq5})) and scaling law (from (\ref{eq8}) to (\ref{eq9}))
driven by 0-boundary condition for the probability distribution
functions for pile-ups $f\left( {n,t} \right)$ for high values of $n \gg n_0$:

\begin{itemize}
\item from ``infinite'' diffusive scaling law for $t<t_{0,p}$
\begin{equation}
\label{eq8}
f_{p,t<t_{0,p} } ( n,\lambda t) \to \lambda ^{-1 \mathord{\left/
{\vphantom {1 2}} \right. \kern-\nulldelimiterspace} 2}f_{p,t<t_{0,p} }\left( {\lambda ^{-1
\mathord{\left/ {\vphantom {1 2}} \right. \kern-\nulldelimiterspace}
2}n,\;t} \right),
\end{equation}

\item to ``semi-infinite'' diffusive scaling law for $t>t_{0,p}$
\begin{equation}
\label{eq9}
f_{p,t>t_{0,p} } \left( {n,\lambda t} \right)\to \lambda ^{-1}f_{p,t>t_{0,p} }\left( {\lambda
^{-1 \mathord{\left/ {\vphantom {1 2}} \right. \kern-\nulldelimiterspace}
2}n,\;t} \right).
\end{equation}
\end{itemize}

\begin{figure}[htbp]
\centerline{
\includegraphics[width=2.8in,height=2.1in]{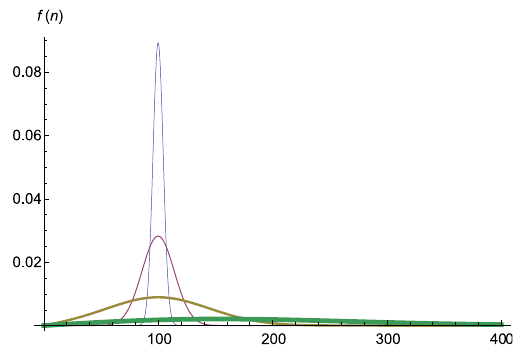}
\includegraphics[width=2.8in,height=2.1in]{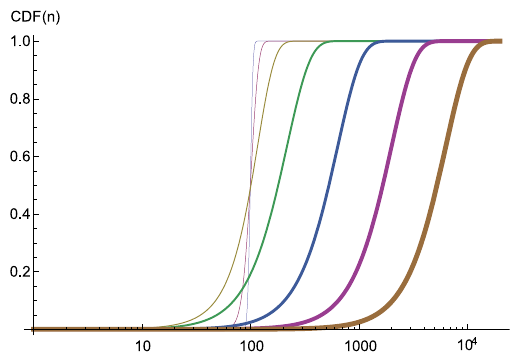}
}
\centerline{a) \hspace{7cm} b)}
\caption{Pile-up size distributions: probability density functions $f(n)$ (a) and their cumulative distribution functions CDF($n$) (b)
for different stages of aggregation kinetics (\ref{eq4_heat}). Thickness of
lines corresponds to the increasing values of $t$ in the order from the thinnest to thickest line:
(a) 100; 1000; 2000; 4000; and (b) 10; 10$^{2}$; 10$^{3}$;
10$^{4}$; 10$^{5}$; 10$^{6}$; 10$^{7}$.}
\label{Pileup_PDFCDF}
\end{figure}

From practical point of view it is not easy to distinguish scaling laws (\ref{eq8}) and (\ref{eq9})
and difference between them by analysis of probability density functions (PDFs) $f(n)$
for various moments of time (Fig.~\ref{Pileup_PDFCDF}a),
especially from scarce and noisy experimental data.
But it is much more easier to observe the scaling laws by analysis of
the cumulative distribution functions.
For example, the visual comprehension of scaling law (\ref{eq9}) can be easily obtained by
viewing the steady shift of the correspondent cumulative
distribution functions (CDFs)  to the right side in Fig.~\ref{Pileup_PDFCDF}b.
The shift becomes diffusive (i.e. proportional to $\sqrt{t}$) after $t>t_{0,p}=10^4$,
when the first cluster will reach the 0-boundary,
i.e. by continuous detachments some cluster will decrease its size to 0 and disappear.

It means that one can determine the qualitatively different scaling regimes
--- ``infinite'' (\ref{eq8}) and ``semi-infinite'' (\ref{eq9}) --- in this aggregation model with
additive rules by the scaling analysis of the probability distribution
function $f\left( {n,t} \right)$ for high values of $n \gg n_0 $.
As it will be shown below in section \ref{section:scaling}, scaling analysis
of cumulative density functions is more efficient for practical purposes.
The unbounded
case (Cauchy problem for infinite domain $-\infty <n<\infty )$ corresponds
to a situation with the constant number of clusters (monomers only
redistribute between them), and the bounded one (first boundary value
problem for domain $0\le n<\infty )$~--- to a situation with irreversible decrease of
the number of clusters.

\paragraph{Wall - maximum active surface}
For the case $\alpha = 1$ clusters have the maximum active
surface and under condition $K_d(n) = K_a(n)$ one can get:

\begin{equation}
\label{eq10}
\frac{\partial f_w\left( {n,t} \right)}{\partial t}=a\frac{\partial ^2\left(
{n\;f_w\left( {n,t} \right)} \right)}{\partial n^2},
\end{equation}

which has numerous particular solutions $f\left( {n,t} \right)\sim F\left(
{n \mathord{\left/ {\vphantom {n t}} \right. \kern-\nulldelimiterspace} t}
\right)$ that are dependent on the initial and boundary conditions. Again
for illustrative purpose we consider the time evolution of the initial
singular distribution of clusters of the same size ($n_0 )$ that exchange
the monomers. It is actually the first boundary value problem for domain
$0\le n<\infty $, $f\left( {0,t} \right)=0$, $f\left( {n,0} \right)=\delta
\left( {n-n_0 } \right)$ with the following solution:

\begin{equation}
\label{eq11}
f_w\left( {n,t} \right)=\frac{\sqrt {n_0 } }{at\sqrt n }\exp \left[
{-\frac{\left( {n+n_0 } \right)}{at}} \right]I_1 \left( {\frac{2\sqrt {nn_0
} }{at}} \right).
\end{equation}

Again, on the initial stage of this one-step aggregation process the
probability distribution function $f\left( {n,t} \right)$ does not ``feel''
the boundary condition $f\left( {0,t} \right)=0$, because from the physical
point of view the cluster size cannot reach the boundary immediately, but
only after some time $t_{0,w}$, when the first cluster disappear due to
detaching monomers. When $t \ll t_{0,w} $ all clusters have nearly the same size $n \approx n_0$, 
and the solution of (\ref{eq10})
has not influenced by differences in $n$ significantly and that is why it has
approximate solution, which is close to solution of the heat equation (\ref{eq7}):

\begin{equation}
\label{eq12}
f_{w,t<t_{0,w} } \left( {n,t} \right)\to \frac{1}{2\sqrt {\pi at} }\exp \left[
{-\frac{(n-n_0)^2}{4at}} \right].
\end{equation}

Because the Bessel function $I_1(z) \rightarrow z/2$ for $z \rightarrow 0$
for the later stage $t \gg t_{0,w}= 2\sqrt{n n_0}/(at)$ and for $n \gg n_0$ solution (\ref{eq11})  of (\ref{eq10}) will be close to:

\begin{equation}
\label{eq13}
f_{w,t>t_{0,w} } \left( {n,t} \right)\to \frac{n_0 }{a^2t^2}\exp
\left[ {-\frac{n}{at}} \right].
\end{equation}

Thus, again in the same process
we have change of size distribution (from (\ref{eq12}) to (\ref{eq13})) and scaling law (from (\ref{eq14}) to (\ref{eq15}))
driven by 0-boundary condition for the probability distribution
function of walls $f\left( {n,t} \right)$ for $n \gg n_0 $:

\begin{itemize}
\item from ``infinite'' diffusive scaling law for $t<t_{0,w}$
\begin{equation}
\label{eq14}
f_{w,t<t_{0,w} } \left( {n,\lambda t} \right)\to \lambda ^{-1 \mathord{\left/
{\vphantom {1 2}} \right. \kern-\nulldelimiterspace} 2}f_{w,t<t_{0,w} }\left( {\lambda ^{-1
\mathord{\left/ {\vphantom {1 2}} \right. \kern-\nulldelimiterspace}
2}n,\;t} \right),
\end{equation}

\item to ``semi-infinite'' ballistic (linear) scaling law for $t>t_{0,w}$ and $n \gg n_0$
\begin{equation}
\label{eq15}
f_{w,t>t_{0,w} } \left( {n,\lambda t} \right)\to \lambda ^{-2}f_{w,t>t_{0,w} }\left( {\lambda^{-1} n,\; t} \right).
\end{equation}
\end{itemize}

\begin{figure}[htbp]
\centerline{
\includegraphics[width=2.8in,height=2.1in]{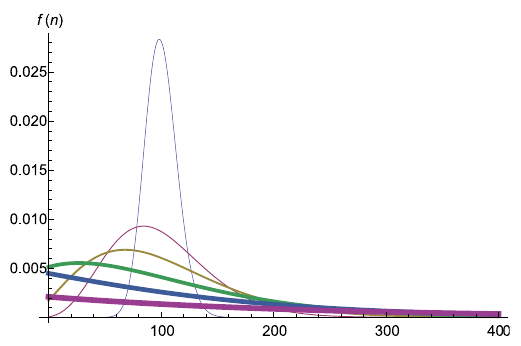}
\includegraphics[width=2.8in,height=2.1in]{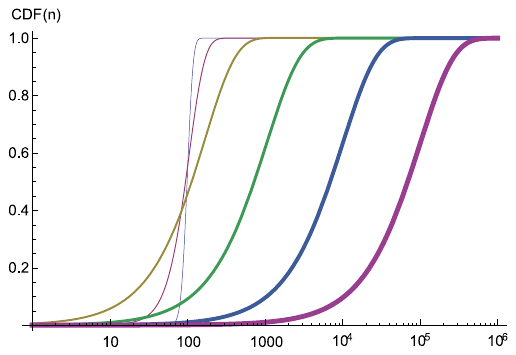}
}
\centerline{a) \hspace{7cm} b)}
\caption{Wall size distributions for different stages of aggregation kinetics
(a) and their cumulative distribution functions (b). Thickness of lines
corresponds to the values of $t$ in the order from the thickest to thinnest
line: (a) 1; 10; 20; 40; 80; 160; and (b) 1; 10; 10$^{2}$; 10$^{3}$; 10$^{4}$;
10$^{5}$}
\label{Wall_PDFCDF}
\end{figure}

Again, as in the case with pile-ups, it is not easy to distinguish scaling laws (\ref{eq14}) and (\ref{eq15})
and difference between them by analysis of PDFs $f(n)$
for various times (Fig.~\ref{Wall_PDFCDF}a), especially from experimental data.

And it is much more easier to observe the scaling laws by analysis of
the cumulative distribution functions (Fig.~\ref{Wall_PDFCDF}b).
The visual comprehension of scaling law (\ref{eq15}) can be easily obtained by
viewing the steady shift of the cumulative
distribution functions (CDFs) to the right side in Fig.~\ref{Wall_PDFCDF}b.
The shift becomes linear (i.e. proportional to $t$)
after $t>t_{0,w}=10^2$ (but not after $t>t_{0,w}=10^4$, like it was for pile-ups!),
when the first cluster will reach the 0-boundary,
i.e. when some cluster by continuous detachments will decrease its size to 0 and disappear.
But the most essential point is that shortly after $t>t_{0,w}$ the size distribution for
wall configuration becomes exponential (\ref{eq13}), that is scale-free one and without
any ``apparent'' peak value.

The assumption as to the better scaling analysis of CFDs in comparison to PDFs
is related to the following simple property:
if PDF $f(n,t)$ is scaled by some law like
$f(n,\lambda t) = \lambda^\mu f( \lambda^\nu x, t)$,
than its CDF $g(n,t)$ is scaled by the following law:

\begin{eqnarray}
\label{eq16}
g(n,\lambda t) 
= 
\frac{ \int_0^n f(x,\lambda t) dx }{\int_0^\infty f(x, \lambda t) dx}
\to
\frac{ \int_0^n \lambda^\mu f( \lambda^\nu x, t) d( \lambda^\nu x) }{\int_0^\infty \lambda^\mu f( \lambda^\nu x, t) d( \lambda^\nu x)} \nonumber
\\
\to [y = \lambda^\nu x] 
\to \frac{ \int_0^{\lambda^\nu n} f(y, t) dy }{\int_0^\infty f( y, t) dy}
= g(\lambda^\nu n, t).
\end{eqnarray}

This means that if PDF $f(n,t)$ can be scaled by stretching its ordinates ($\lambda^\mu$) and abscissas ($\lambda^\nu$),
then its CDFs $g(n,t)$ can be scaled by stretching abscissas ($\nu\neq0$) only, but not by ordinates ($\mu=0$),
which will be preserved.

Thus, change of scaling for pile-up CDF $g_p\left( {n,t} \right)$ should be:

\begin{itemize}
\item from ``infinite'' diffusive scaling law for $t<t_{0,p}$
\begin{equation}
\label{eq17}
g_{p,t<t_{0,p} } (n,\lambda t) \to g_{p,t<t_{0,p} } (\lambda^{-1/2}n, t),
\end{equation}

\item to ``semi-infinite'' diffusive scaling law for $t>t_{0,p}$
\begin{equation}
\label{eq18}
g_{p,t>t_{0,p} } (n,\lambda t) \to g_{p,t>t_{0,p} } (\lambda^{-1/2}n, t).
\end{equation}
\end{itemize}

Similarly, change of scaling for wall CDF $g_w\left( {n,t} \right)$ should be:

\begin{itemize}
\item for $t<t_{0,w}$ (``infinite'' diffusive scaling law)
\begin{equation}
\label{eq19}
g_{w,t<t_{0,w} } (n,\lambda t) \to g_{w,t<t_{0,w} } (\lambda^{-1/2}n, t),
\end{equation}

\item for $t>t_{0,w}$ and $n \gg n_0$ (``semi-infinite'' ballistic (linear) scaling law)
\begin{equation}
\label{eq20}
g_{w,t>t_{0,w} } (n,\lambda t) \to g_{w,t>t_{0,w} } (\lambda^{-1}n, t),
\end{equation}
\end{itemize}

    In comparison, nonlinear Leyvraz--Redner approach \cite{Leyvraz} on the basis of the linked Smolukhovski nonlinear equations \cite{Smoluchowski1916} becomes linear (under conditions of the product kernel $K(i,j)=(ij)^\lambda$
    and conserved number of all monomers $\sum n f(n,t)=1$)
     after rescaling the time variable by moments of the size distribution $M_\lambda = \sum n^\lambda f(n,t)$, as it was shown in similar Ben-Naim-Krapivsky theory for exchange driven growth \cite{ben2003exchange} and Lin-Ke theory for migration-driven aggregation \cite{ke2002kinetics,lin2003kinetics,lin2005exchange}, which were also formulated on the basis of the linked Smolukhovski nonlinear equations \cite{Smoluchowski1916}. The one-step model proposed here is
     in agreement with results of scaling analysis on the basis of the approximate ansatz function in Ben-Naim-Krapivsky theory for exchange driven growth, namely with Eq.(8) and Eq.(10) in their work \cite{ben2003exchange} taking into account the rescaled time variable.
     Moreover, the rescaled time variable in Eq.(3) in Ref.\cite{ben2003exchange} $\tau = \int_0^t dt' M_\lambda (t')$ expresses the essential difference between the one-step formulation of aggregation process proposed here
     (where \emph{single} transitions allowed and between \emph{adjacent} integers only \cite{vanKampen},
     which manifests itself in \emph{linearity} and $K(i,j)=K(1,j)=j^\lambda$)
     and the multi-step formulation of aggregation process \cite{Leyvraz,ke2002kinetics,lin2003kinetics,lin2005exchange}
     (where \emph{multiple} transitions allowed and between \emph{any} integers,
     which manifests itself in \emph{non-linearity}, $K(i,j)=(ij)^\lambda$,
     and time dependence on the available cumulative active surface expressed by multiplier $M_\lambda$ in Eq.(3) in Ref.\cite{ben2003exchange}).

One of drawbacks of the simplified model is that the equations (\ref{eq4_heat}) and (\ref{eq10})
and their solutions (\ref{eq5}) and (\ref{eq11}) give
the idealized and rough representation of the aggregation kinetics.
In fact, solutions (\ref{eq5}) and (\ref{eq11}) allows the ``long-tails'' of distribution,
i.e. non-zero density of clusters for arbitrary high or low values of $n$
in the ranges $n>n_0$ and $n<n_0$ for any stage of evolution.
For example, any initial cluster of size $n=10^2$
will need at least $t=10^2$ detachment steps to disappear.
But the analytical solution (5) gives the non-zero values of cluster distribution
$f(n,t)$ for any small $n$, i.e. does not prohibit immediate disappearance of this initial cluster
even at the first step after start of the aggregation process.
In a real physical process it is impossible due to physical limitations of
diffusive character of the one-step aggregation process among clusters
with pile-up morphology.
This situation is well-known and actively investigated in the class of so-called
"first-passage problems" \cite{vanKampen}.
That is why, the analytical solutions and predictions of the model cannot be directly applied
to the correspondent practical situations
without taking into account the limits of their physical counterparts.
Below, the aforementioned change of size distribution and scaling law driven by 0-boundary condition is investigated by the following simulations.

\section{Simulation}
The two aforementioned primitive cases of cluster aggregation were
simulated by Monte Carlo method to illustrate the different cluster distributions in
different aggregation kinetics. The numerous initial configurations of
clusters with various numbers of monomers in each of them were used in simulations
with a conserved number of $10^6$ aggregating monomers as a whole.
Below the results on initial configurations of $10^4$
clusters with $10^2$ monomers are considered (and the other vast configurations
will be reported later \cite{gordienkoIJMPB2012}).
Further the analytical calculations in section \ref{Model} will be compared
with the results of simulation and for this purpose
the number of Monte Carlo sweeps (MCSs) will be assumed to equal to the number of time steps $t$.
All simulation runs were performed in the distributed computing
infrastructure (DCI) ``SLinCA@Home'' (Scaling Law in Cluster Aggregation) (http://dg.imp.kiev.ua/slinca)
on the basis of BOINC SZTAKI Desktop Grid (DG) technology \cite{Kacsuk,Urbach}
through the linked science gateway portal on the basis of WS-PGRADE technology \cite{WSPGRADE2012}.

\paragraph{Pile-up --- minimum active surface}

\begin{figure}[t!]
\centerline{
\includegraphics[width=7cm,height=7cm]{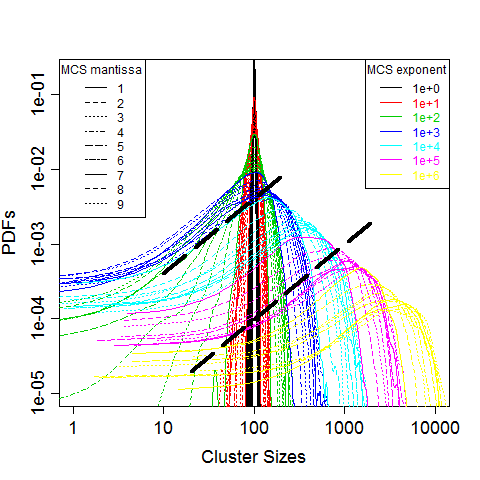}
\hspace{.1cm}
\includegraphics[width=7cm,height=7cm]{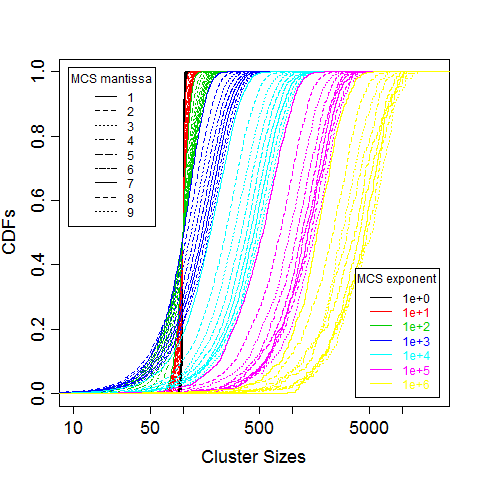}
}
\centerline{a) \hspace{7cm} b)}
%
%
\caption{Size distributions of \textit{pile-ups} for different stages of aggregation kinetics: (a) PDFs in double logarithmic coordinates, and (b) CDFs in logarithmic-linear coordinates. Here and below figures in legends denote the number of Monte Carlo sweeps (MCSs) represented 
in decimal logarithmic notation with the mantissas and exponents noted in the legends.}
\label{Pileup_PDFCDF_simulation}
\end{figure}

The ``pile-up'' aggregation of monomers (see
Fig.~\ref{Examples}a) has a minimum active surface and it was simulated
for unbiased migration-driven aggregation.
The kinetics of rearrangements from the
initial configuration is shown in Fig.\ref{Pileup_PDFCDF_simulation}.
After some MCSs of the initial configuration the broad peak appears, which
corresponds to the average cluster size. The size distributions of pile-ups
%
(Fig.~\ref{Pileup_PDFCDF_simulation}a)
become asymmetric, but they preserve their one-peak shapes.
And their CDFs
%
(Fig.~\ref{Pileup_PDFCDF_simulation}b)
demonstrate the ``diffusive'' scaling law,
which becomes evident after some number of MCSs, namely after $t>t_{0,p}=10^4$.
It can be seen visually by the shift of the simulated CDFs to the right side
(proportionally to $\sqrt{t}$) on
%
the log-linear plot in 
%
(Fig.~\ref{Pileup_PDFCDF_simulation}b)
and can be compared with the same shift of the analytically calculated CDFs in Fig.~\ref{Wall_PDFCDF}b.
The two straight thick dash lines are given in
%
Fig.~\ref{Pileup_PDFCDF_simulation}a
as a guide for eyes to emphasize that $f(n,t) \sim n$  for $n<n_0 $ and $t>t_{0,p}$, that is follows from (\ref{eq6}).

\paragraph{Wall --- maximum active surface}

\begin{figure}[htbp]
\centerline{
\includegraphics[width=7cm,height=7cm]{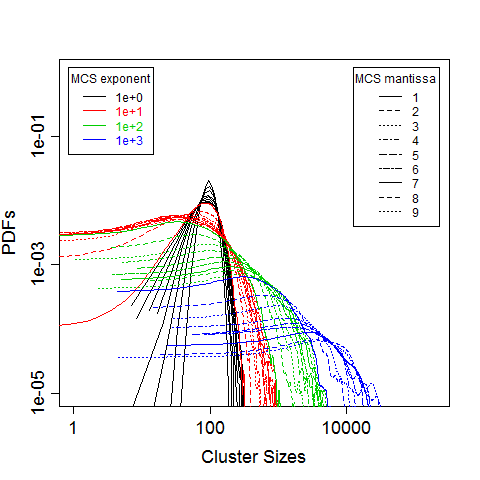}
\hspace{.1cm}
\includegraphics[width=7cm,height=7cm]{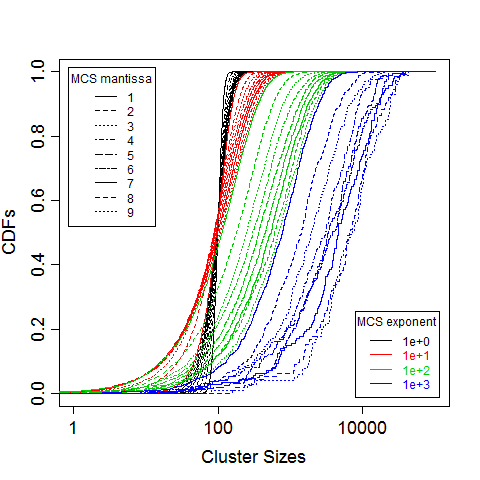}
}
\centerline{a) \hspace{7cm} b)}
%
%
\caption{Size distributions of \textit{walls} for different stages of aggregation kinetics: (a) PDFs in double logarithmic coordinates, and (b) CDFs in logarithmic-linear coordinates.}
\label{Wall_PDFCDF_simulation}
\end{figure}

%
The ``wall'' aggregation of monomers (see
Fig.~\ref{Examples}b) has a maximum active surface and it was simulated
for unbiased migration-driven aggregation.
The kinetics of rearrangements from initial configurations is
shown in Fig.~\ref{Wall_PDFCDF_simulation}.
It can be seen that size distributions of walls
(Fig.~\ref{Wall_PDFCDF_simulation}a,b)
become asymmetric also. Moreover, they do \textit{not} preserve their one-peak shapes after some number of
MCSs and become scale-free, as it was predicted by the analytical solution (\ref{eq13}) for $t>t_{0,w}$.
After some Monte Carlo sweeps (MCSs) the initial peak, that
corresponds to the initial cluster size $n_0$, disappear and the scale-free
distribution \textit{without distinctive peak} appears.
This transition is much more evident in the log-log representation of cluster size distributions on
%
Fig.~\ref{Wall_PDFCDF_simulation}a,
where availability of peak (after $t>10^1$) is apparent,
such ``peak'' cannot be distinguished from error limits,
and its position is highly dependent on the selected bin size.
Their CDFs
%
(Fig.~\ref{Wall_PDFCDF_simulation}b)
also demonstrate the ``semi-infinite'' ballistic (linear) scaling law (\ref{eq15}), which becomes evident after some number of MCSs,
when the shift of CDFs to the right side of the plot in
%
Fig.~\ref{Wall_PDFCDF_simulation}b
becomes linearly proportional to $t$ after $t>t_{0,w}=10^2$.
One can easily compare it with the similar shift of CDFs in Fig.~\ref{Wall_PDFCDF}b).

    To avoid the subjective estimations (like visual examination) of PDF/CDF curves, the simulation results
    were analyzed by the scaling, fitting, moment, and bootstrapping analysis in the next sections.

%
\section{Change of scaling and distribution type}
\subsection{Scaling analysis}\label{section:scaling}
Comparison of analytical and simulation results shows that monomer
aggregation (at least, in pile-up and wall configurations of clusters) can
be satisfactorily described by the proposed theoretical model of one-step
aggregation processes.
The important thing is that some scaling laws can be determined
experimentally by scaling analysis of size distributions
%
(PDFs and CDFs)
of the real aggregating ensembles. Below some attempts of such analysis 
for 63 different pile-up PDF/CDF pairs and
for 36 different wall PDF/CDF pairs 
are shown on the basis of aforementioned simulation results
and with usage of R language and environment for statistical computing \cite{R2008}.

\paragraph{Pile-up --- minimum active surface}

\begin{figure}[htbp]
\centerline{
\includegraphics[width=7cm,height=7cm]{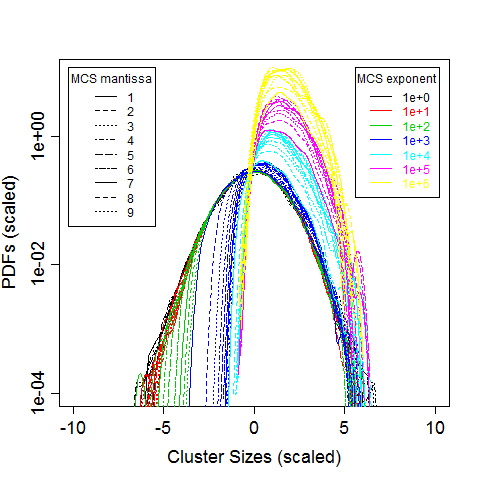}
\hspace{.1cm}
\includegraphics[width=7cm,height=7cm]{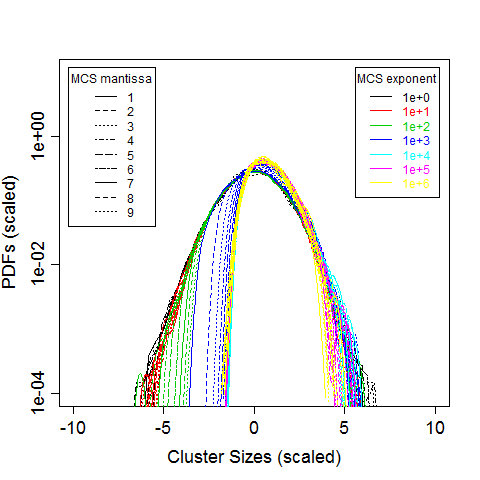}
}
\centerline{a) \hspace{7cm} b)}
\centerline{
\includegraphics[width=7cm,height=7cm]{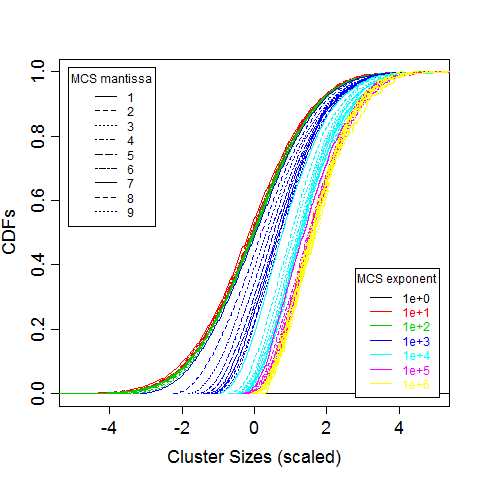}
\hspace{.1cm}
\includegraphics[width=7cm,height=7cm]{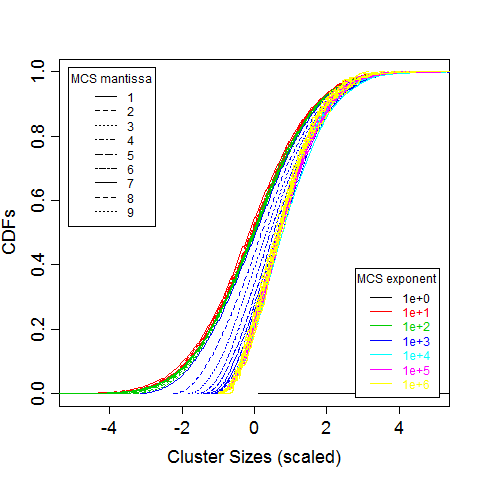}
}
\centerline{c) \hspace{7cm} d)}
%
%
\caption{Change of scaling in aggregation kinetics of \textit{pile-ups} presented by differently scaled
PDFs (from Fig.~\ref{Pileup_PDFCDF_simulation}a): 
(a) scaling laws (\ref{eq8}) for $t<t_{0,p}=10^4$ and (\ref{eq9}) for $t>t_{0,p}=10^4$, 
(b) scaling laws (\ref{eq8}) for $t<t_{0,p}=10^4$ and (\ref{eq21}) for $t>t_{0,p}=10^4$;
and CDFs (from Fig.~\ref{Pileup_PDFCDF_simulation}b): 
(c) scaling laws (\ref{eq17}) for $t<t_{0,p}=10^4$ and (\ref{eq18}) for $t>t_{0,p}=10^4$, 
(d) scaling laws (\ref{eq17}) for $t<t_{0,p}=10^4$ and (\ref{eq21}) for $t>t_{0,p}=10^4$.}
\label{Pileup_PDFCDF_scaled}
\end{figure}

\begin{figure}[!h]
\centerline{
\includegraphics[width=7cm,height=7cm]{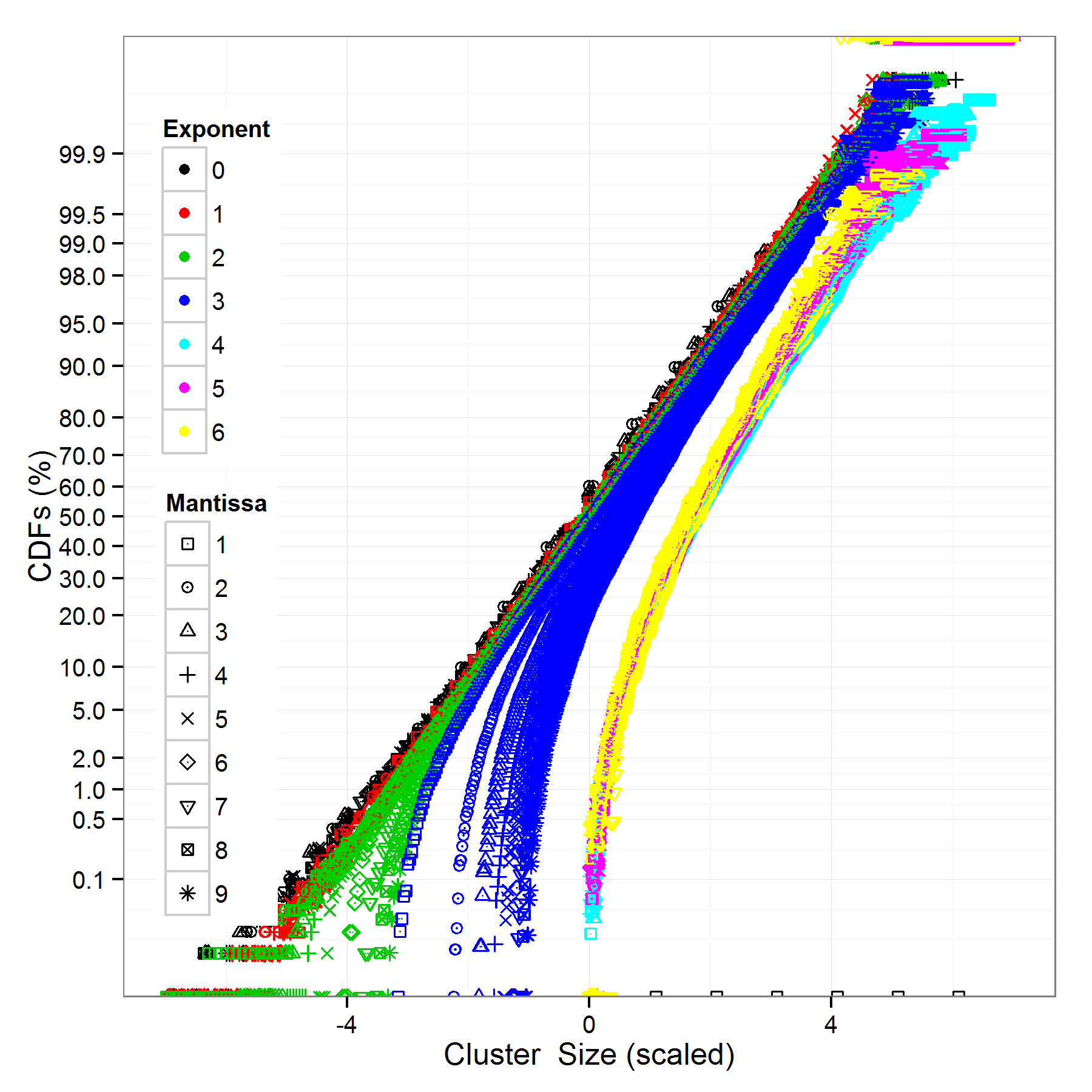}
\hspace{.1cm}
\includegraphics[width=7cm,height=7cm]{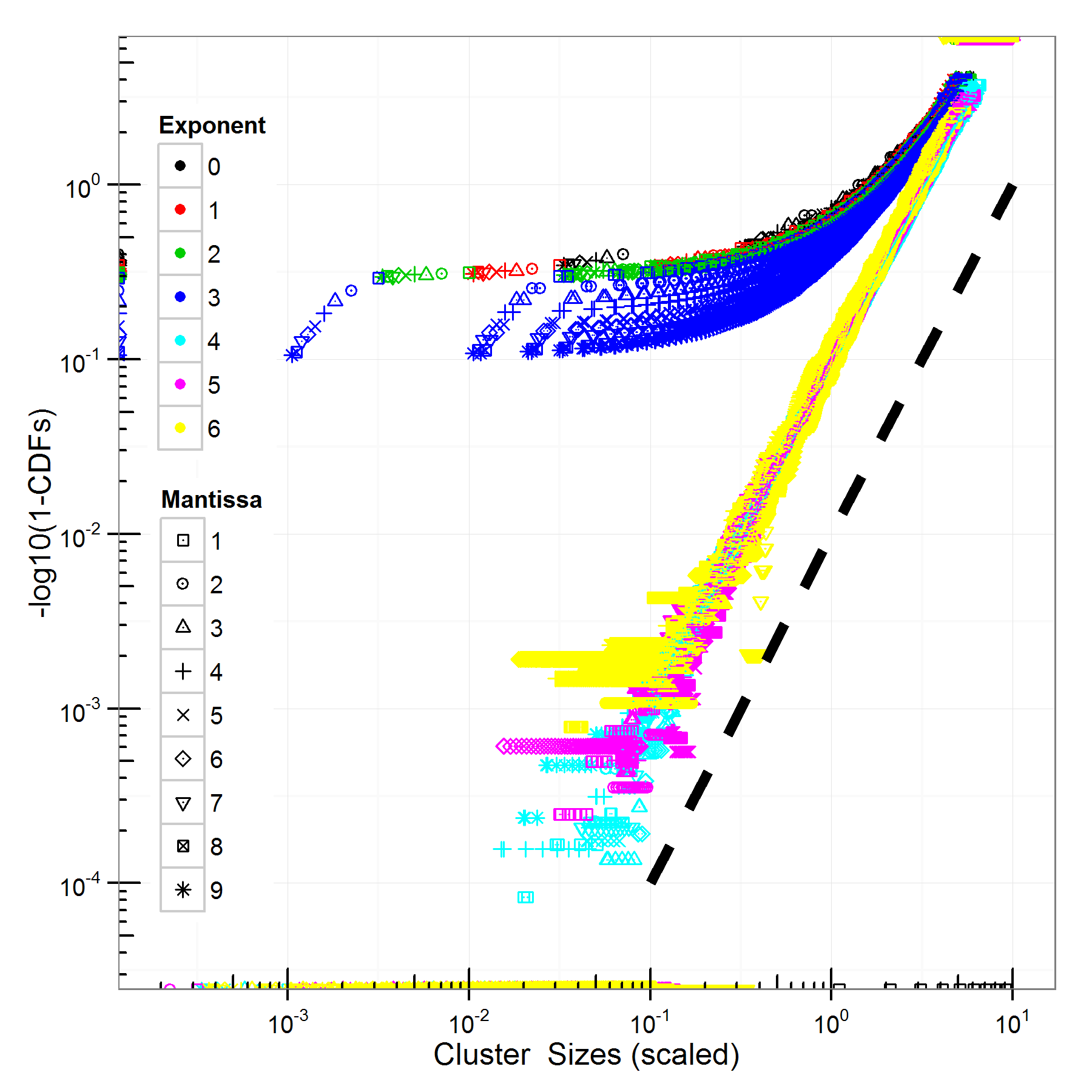}
}
\centerline{a) \hspace{7cm} b)}
%
%
\caption{Different scaling laws in aggregation kinetics of \textit{pile-ups} presented 
for scaled CDFs ($g_p$) (from Fig.~\ref{Pileup_PDFCDF_simulation}b and Fig.~\ref{Pileup_PDFCDF_scaled}c) 
on the plots with differently scaled ordinate axis: 
(a) on the probability plot, where CDFs collapsing to straight line (i.e. to the normal distribution), and 
(b) on the Weibull-scaled plot, where CDFs collapsing to straight line (i.e. to the Weibull distribution, see details in section \ref{section:fitting}).}
\label{Pileup_CDF_Gauss_vs_Weibull}
\end{figure}

%
Application of two scaling laws (``infinite'' (\ref{eq8}) for $t<t_{0,p}=10^4$ and ``semi-infinite'' (\ref{eq9}) for $t > t_{0,p}=10^4$)
to simulated PDFs (Fig.~\ref{Pileup_PDFCDF_scaled}a),
and scaling laws (``infinite'' (\ref{eq17}) for $t<t_{0,p}=10^4$ and ``semi-infinite'' (\ref{eq18}) for $t>t_{0,p}=10^4$) --- to simulated CDFs (Fig.~\ref{Pileup_PDFCDF_scaled}c) of pile-ups
shows the clear difference between these two scaling regimes.
%
%
The ``infinite'' diffusive scaling law (\ref{eq8}) corresponds to the initial stage
(in the range of MCS steps $t<t_{0,p}=10^4$),
i.e. the 18 scaled PDFs (Fig.~\ref{Pileup_PDFCDF_scaled}a) and CDFs (Fig.~\ref{Pileup_PDFCDF_scaled}c) curves
collapse to the initial \emph{symmetric} size distribution.
Some inclinations from perfect collapse are related to influence of the boundary condition
and slow transition to the different scaling law (for the left tail),
and worse precision, because of the low number of biggest cluster (for the right tail).
The same collapse of the scaled CDF curves to the \textit{initial} straight line can be seen
on the probability plot (Fig.~\ref{Pileup_CDF_Gauss_vs_Weibull}a).
The straight dashed line on the probability plot corresponds to CDF of normal distribution,
i.e. this collapse demonstrates good correspondence to solution (\ref{eq7}).

The ``semi-infinite'' diffusive scaling law (\ref{eq9}) takes place later (in the range of MCS steps $t>t_{0,p}=10^4$),
i.e. the $>30$ scaled PDFs (Fig.~\ref{Pileup_PDFCDF_scaled}a) and CDFs (Fig.~\ref{Pileup_PDFCDF_scaled}c) curves
collapse to the asymptotic \emph{asymmetric} size distribution.
The same collapse of the scaled CDF curves to the \textit{asymptotic} straight line can be seen
on the plot (Fig.~\ref{Pileup_CDF_Gauss_vs_Weibull}b) with the scaled ordinate axis (see the label near this axis).
The straight dashed line on such plot corresponds to CDF of Weibull distribution,
which slope ($\approx 2$ in Fig.~\ref{Pileup_CDF_Gauss_vs_Weibull}b)
should correspond to the power of $n$ under exponent in (\ref{eq6})
(see the details in \cite{gordienkoIJMPB2012}, namely, Eq.(5) and Eq.(6)).
This collapse demonstrates good correspondence to solution (\ref{eq6}),
and the detailed comparison of Weibull distribution and solution (\ref{eq6}) will be given below
in section \ref{section:fitting}.

The approximate transition region ($10^3<t_b<10^4$) between these two scaling laws
can be observed in Fig.~\ref{Pileup_PDFCDF_scaled}a,c and Fig.~\ref{Pileup_CDF_Gauss_vs_Weibull}
as the 15-20 shifting to right curves (denoted by blue color in electronic version),
i.e. near the moment of the first-passage of a cluster through the point of disappearance ($n=0$).
The matter is after reaching the 0-boundary ($10^3<t_b<10^4$)
the number of clusters decreases and the mean of the cluster distribution increases like $\left< n \right> \sim \sqrt{t}$.
Thus, the PDFs and CDFs 
(mostly corresponding to solution (\ref{eq5}) with taking into account $n_0$)
can be scaled to the better collapse near peaks (Fig.~\ref{Pileup_PDFCDF_scaled}b,d)
by the other ``diffusive scaling law with moving mean'':

\begin{eqnarray}
    \label{eq21}
    f_{p,t\approx t_b } (n,\lambda t) \to \lambda^{-1/2} f_{p,t\approx t_b } (\lambda^{-1/2} (n-\lambda^{1/2}),t ), \nonumber
    \\
    g_{p,t\approx t_b } (n,\lambda t) \to g_{p,t\approx t_b } (\lambda^{-1/2} (n-\lambda^{1/2}),t ).
\end{eqnarray}

It should be emphasized that this ``diffusive scaling law with moving mean'' is good for the regions near peak of distribution
($n \approx n_0$) and bad for the right tail ($n \gg n_0$) (see Fig.~\ref{Pileup_PDFCDF_scaled}b).

\paragraph{Wall --- maximum active surface}

\begin{figure}[htbp]
\centerline{
\includegraphics[width=7cm,height=7cm]{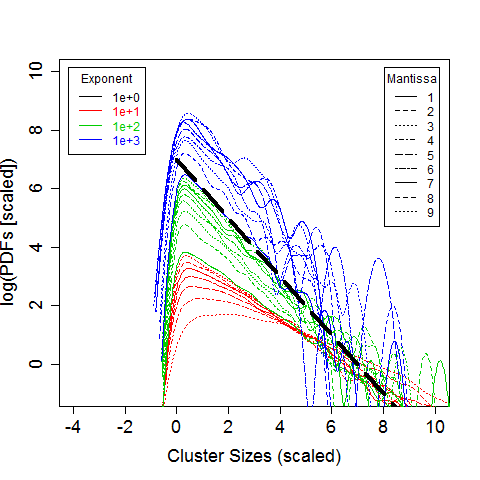}
\hspace{.1cm}
\includegraphics[width=7cm,height=7cm]{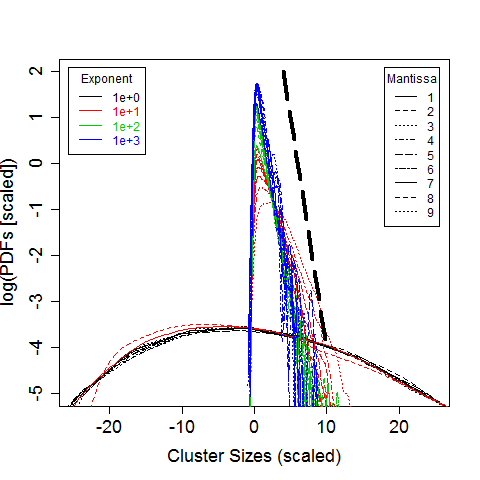}
}
\centerline{a) \hspace{7cm} b)}
%
%
\caption{Change of scaling in aggregation kinetics of \textit{walls} represented by scaled: (a) PDFs, and (b) CDFs.}
\label{Wall_PDF_scaled}
\end{figure}

\begin{figure}[htbp]
\centerline{
\includegraphics[width=7cm,height=7cm]{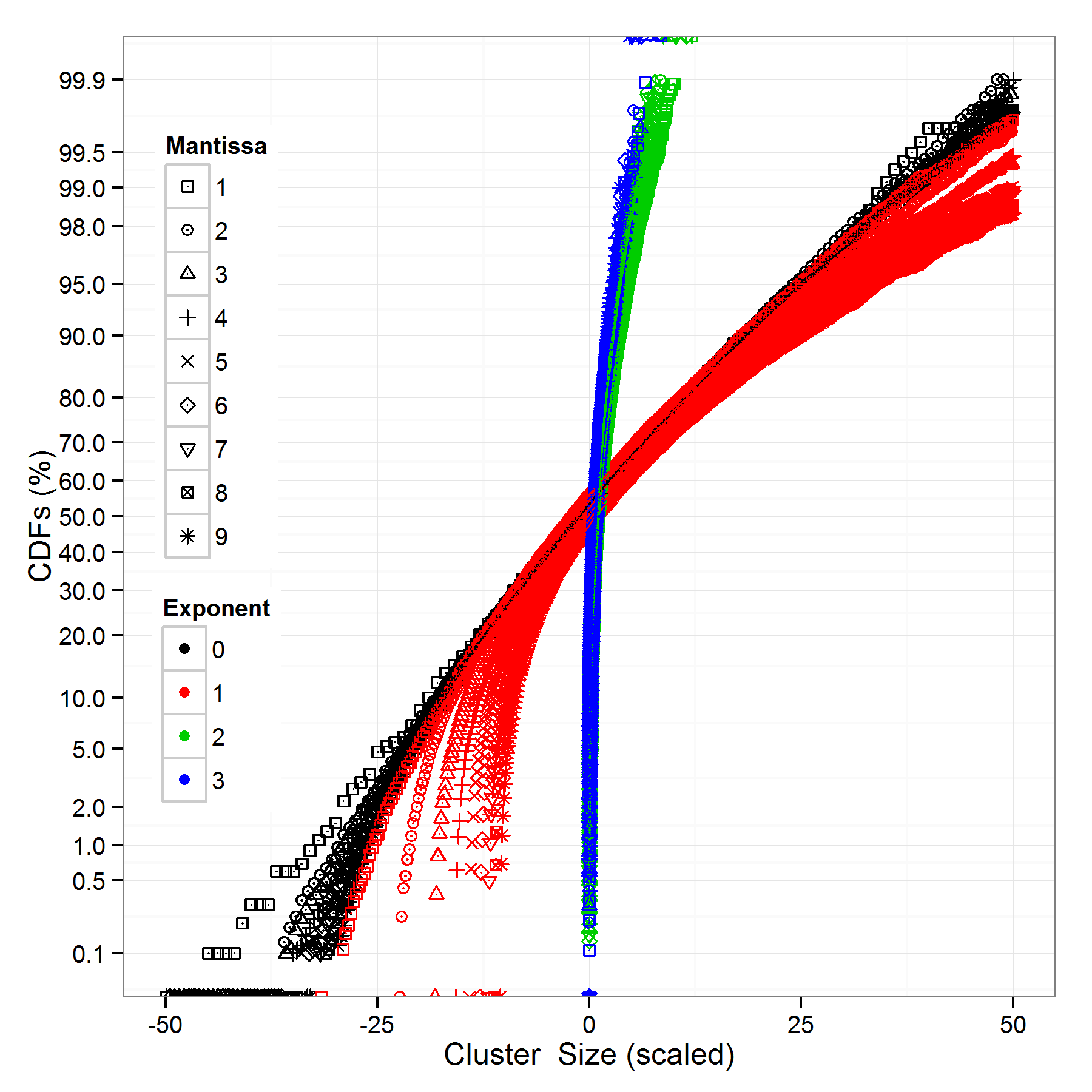}
\hspace{.1cm}
\includegraphics[width=7cm,height=7cm]{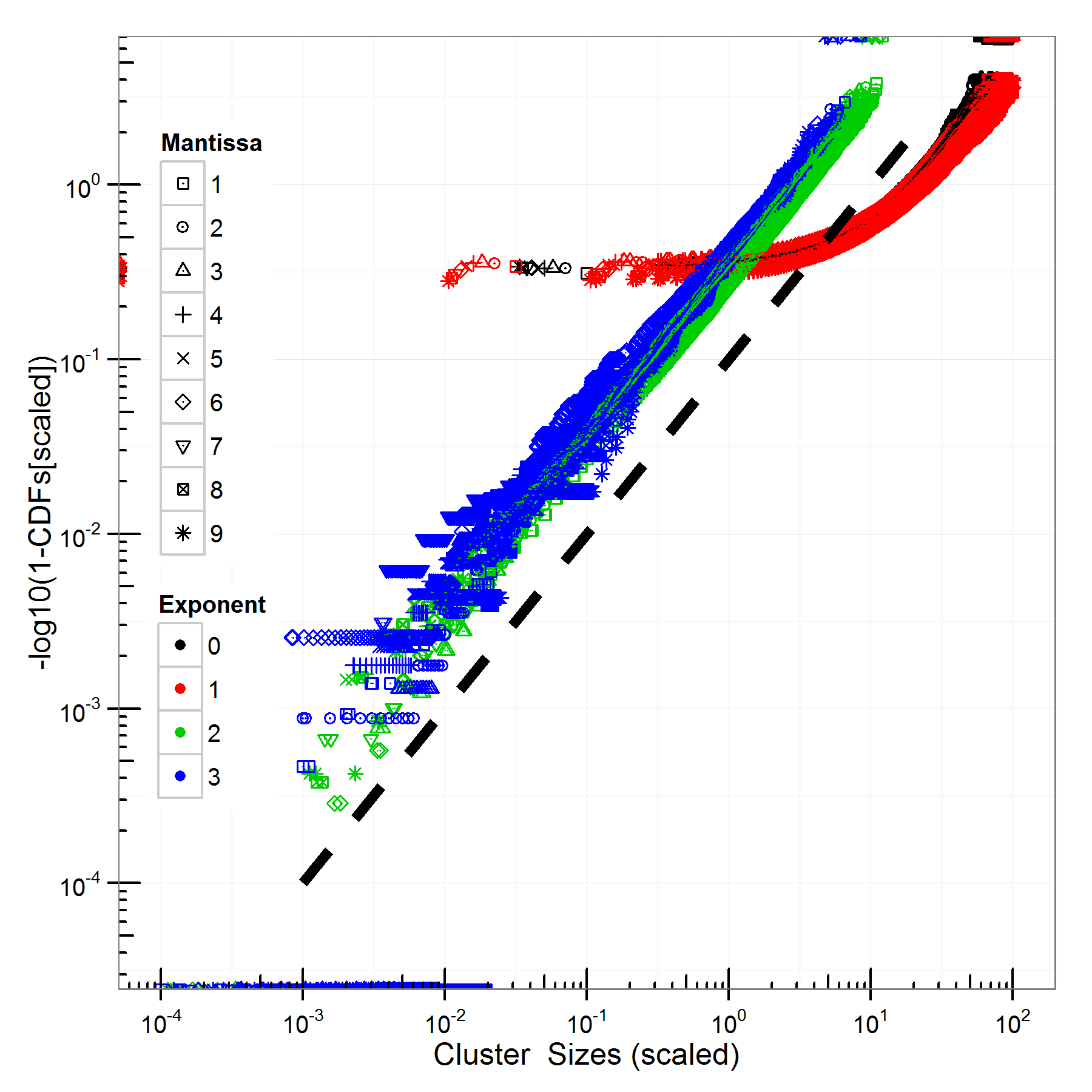}
}
\centerline{a) \hspace{7cm} b)}
%
%
\caption{Different scaling laws in aggregation kinetics of \textit{walls} presented
for scaled CDFs ($g_p$) (from Fig.~\ref{Wall_PDFCDF_simulation}b)
on the plots with differently scaled ordinate axis:
(a) on the probability plot, where CDFs collapsing to nearly straight line (i.e. close to the normal distribution), and
(b) on the Weibull-scaled plot, where CDFs collapsing to straight line (i.e. to the Weibull distribution, 
see details in section \ref{section:fitting}).}

\label{Wall_CDF_Gauss_vs_Weibull}
\end{figure}

Similarly, application of of two scaling laws (``infinite'' diffusive (\ref{eq14}) for $t<t_{0,w}=10^2$
and ``semi-infinite'' ballistic (linear) (\ref{eq15}) for $t > t_{0,w}=10^2)$
to simulated PDFs (Fig.~\ref{Wall_PDF_scaled}) and CDFs (Fig.~\ref{Wall_CDF_Gauss_vs_Weibull}) of walls
shows the other clear difference between two scaling regimes.
The ``infinite'' diffusive scaling law (\ref{eq14}) corresponds to the initial stage $t<t_{0,w}$,
i.e. the 9 scaled PDFs collapse to the initial symmetric size distribution in Fig.~\ref{Wall_PDF_scaled}b;
and the 9 scaled CDFs --- to the initial \textit{straight} line on the probability plot in Fig.~\ref{Wall_CDF_Gauss_vs_Weibull}a.
The ``semi-infinite'' ballistic (linear) scaling law (\ref{eq15}) takes place later (for $t>t_{0,w})$ after some transient period.
The 18 scaled PDFs are attracted to the other \textit{asymmetric} and \textit{scale-free} size distribution,
which is actually the exponential one (see tendency for the right tails in Fig.~\ref{Wall_PDF_scaled});
and the 18 scaled CDFs --- to the asymptotic \textit{straight} line
on the plot (Fig.~\ref{Wall_CDF_Gauss_vs_Weibull}b) with the Weibull-scaled ordinate axis (see the label near this axis).
The straight dashed lines in Fig.~\ref{Wall_PDF_scaled} corresponds to PDF of exponential distribution,
which slope ($\approx 1$ in Fig.~\ref{Wall_PDF_scaled})
should correspond to the power of $n$ under exponent in (\ref{eq13}).

It should be noted that scaling laws (\ref{eq14}) and (\ref{eq15}) are not universal for all $t$ and $n$,
because they were derived from the approximate solutions (\ref{eq12}) and (\ref{eq13}).
Moreover, the exact solution (\ref{eq11}) of equation (\ref{eq10}) contains several combinations of $n$ and $t$
(like $t\sqrt n$, $(n+n_0)t$, $\sqrt n /t$), that will not allow for simple scaling transformation for all $t$ and $n$.
As a result some limiting cases can be emphasized, for example,
collapse of scaled curves in Fig.~\ref{Wall_PDF_scaled}a is not good enough,
because for $n \gg n_l = a^2 t^2/(4n_0)$ the Bessel function $I_1(z) \rightarrow { \exp(z) }/{\sqrt{2 \pi z}}$ for $z \gg 1$
and asymptotic version of solution (\ref{eq16}) will be close to:

\begin{equation}
\label{eq22}
f_{w,n \gg n_l} \left( {n,t} \right) = \frac{ 1 }{2 \sqrt{\pi at} } \frac{ n_0^{1/4} }{ n^{3/4} } \exp(-\frac{n}{at}).
\end{equation}

And PDFs should be better scaled by the other scaling law:
\begin{equation}
    \label{eq23}
    f_{w,n \gg n_l} \left( {n,\lambda t} \right)\to \lambda^{-5/4} f_{w,n \gg n_l}\left( \lambda^{-1} n,t \right).
\end{equation}

The PDF curves scaled by this law (Fig.~\ref{Wall_PDF_scaled}b) demonstrates a good
collapse and correspondence to the asymptotic solution (\ref{eq18}).
It should be noted that PDF is sensitive to the shape (type) of the distribution and CDF is not,
which allows us to find the change of scaling by analysis of CDFs with a higher precision and reliability, than by analysis of PDFs.

Really, the collapse of the scaled CDF curves to the \textit{initial} straight line can be seen
on the probability plot (Fig.~\ref{Wall_CDF_Gauss_vs_Weibull}a).
The straight dashed line on the probability plot corresponds to CDF of normal distribution,
i.e. this not-perfect collapse demonstrates very rough correspondence to the approximate solution (\ref{eq12}) and for the initial stage only ($t<10$).
This unsatisfactory scaling is explained by the appearing difference 
in the cluster sizes, when the initial approximation $n \approx n_0$ becomes not accurate.
That is why (\ref{eq10}) becomes influenced by differences in $n$ significantly and that is why 
its approximate solution (\ref{eq12}), which is close to solution of the heat equation (\ref{eq7})
becomes more inaccurate with time, especially for tails, where difference in $n$ is bigger.
The same collapse of the scaled CDF curves to the \textit{asymptotic} straight line can be seen
on the plot (Fig.~\ref{Wall_CDF_Gauss_vs_Weibull}b) with the Weibull-scaled ordinate axis (see the label near this axis).
The straight dashed line on such plot corresponds to CDF of Weibull (actually exponential one) distribution,
which slope ($\approx 1$ in Fig.~\ref{Wall_CDF_Gauss_vs_Weibull}b)
should correspond to the power of $n$ under exponent in (\ref{eq13}).
This collapse of 20 scaled CDF curves (in the range of 2 decades of MCSs)
demonstrates a good correspondence to solution (\ref{eq13}),
and the detailed comparison of Weibull distribution and solution (\ref{eq13}) will be given below
in section \ref{section:fitting}.

On practice, when the actual analytic solutions, scaling laws, and
availability of such boundary are not known,
such scaling analysis (even for CDFs) can be a very hard and unproductive task.
It is exacerbated by the everlasting problem, because
visual examination of data collapse on the same scaling curve has the subjective nature.
Below, some other useful tools are proposed to determine
the boundary-driven change of scaling and type of distribution.

\subsection{Analysis of fitting distributions}\label{section:fitting}
According to scaling analysis of asymptotic behavior of pile-ups and walls,
their PDF/CDF pairs should change crucially after transition period $t>t_{0,p}$ and $t>t_{0,w}$, respectively. 

\paragraph{Pile-up --- minimum active surface}
In fact, after substitution $\eta_p=2(at)^{1/2}$
(see the details in \cite{gordienkoIJMPB2012}) the PDF from (\ref{eq6}) will be:

\begin{equation}
\label{eq24}
f_p (n)=\frac{2n_0}{\sqrt\pi \eta_p} W(n;\eta_p,\beta_p),
\end{equation}

where $W(n;\eta_p,\beta_p)=(\beta_p/\eta_p)(n/\eta_p)^{\beta_p-1} \exp \left[ -(n/\eta_p)^{\beta_p} \right]$ is a Weibull distribution
(where $\beta_p=2$ is a shape parameter and $\eta_p$ is a scale one), and its CDF will be:

\begin{equation}
\label{eq25}
g_p (n)=1-\exp \left[ -(n/\eta_p)^{\beta_p} \right]=1-\exp \left[ -(n/\eta_p)^2 \right],
\end{equation}

which is the exact CDF for a Weibull distribution with $\beta_p=2$.

\paragraph{Wall --- maximum active surface}
Similarly, after substitution $\eta_w=at$ the PDF from (\ref{eq13}) will be:

\begin{equation}
\label{eq26}
f_w (n)=\frac{n_0 }{\eta_w} W(n;\eta_w,\beta_w),
\end{equation}

where $W(n;\eta_w,\beta_w)==(\beta_w/\eta_w)(n/\eta_w)^{\beta_w-1} \exp \left[ -(n/\eta_w)^{\beta_w} \right]$ is a Weibull distribution,
actually the exponential distribution for $\beta_w=1$
(where $\beta_w=1$ is a shape parameter and $\eta_w$ is a scale one), and its CDF will be:

\begin{equation}
\label{eq27}
g_w (n)=1-\exp \left[ -(n/\eta_w)^{\beta_w} \right]=1-\exp \left[ -n/\eta_w \right],
\end{equation}

which is the exact CDF for a Weibull distribution with $\beta_w=1$.

This means that the change of scaling could be determined by the change:
from the initial normal distribution for pile-ups
(or close to normal for walls) to Weibull distributions 
with the different constant shape parameters $\beta_{p,w}$ 
and the scale parameters $\eta_p=\eta_p(t)$ for pile-ups 
($\eta_w=\eta_w(t)$ for walls), 
which should differently change with time
after transition period $t>t_{0,p}$ for pile-ups (and $t>t_{0,w}$ for walls).

\begin{figure}[htbp]
\centerline{
    \includegraphics[width=7cm,height=7cm]{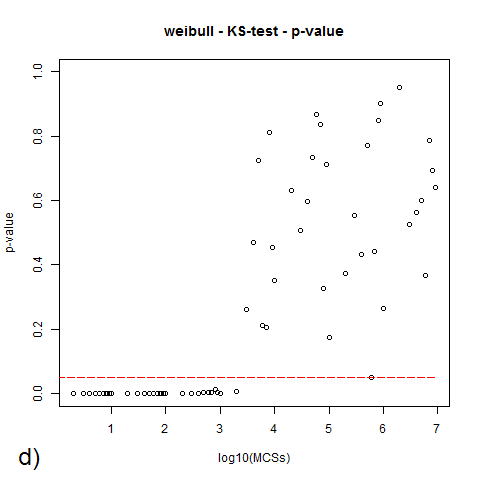}
\hspace{.1cm}
    \includegraphics[width=7cm,height=7cm]{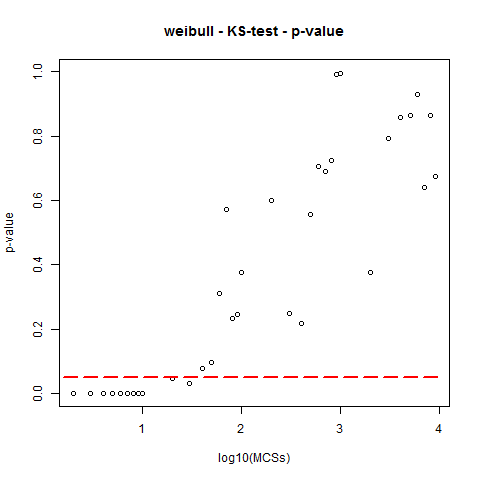}
}
\centerline{a) \hspace{7cm} b)}
\caption{The results (p-values) of the Kolmogorov-Smirnov test for CDFs of: (a)~pile-ups (Fig.\ref{Pileup_PDFCDF_simulation}b), and (b)~walls (Fig.\ref{Wall_PDFCDF_simulation}b).}
\label{Weibull_KS}
\end{figure}

Despite the visually good collapse of CDF curves
in Fig.\ref{Pileup_CDF_Gauss_vs_Weibull}b and Fig.\ref{Wall_CDF_Gauss_vs_Weibull}b,
where the ordinate axes are fitted so, that
the straight dashed lines correspond to Weibull distributions,
the goodness of fits (\ref{eq25}) and (\ref{eq27}) was checked in the following statistical test.
The null and alternative hypotheses were:
\begin{itemize}
  \item $H_0$: Data come from the Weibull distributions,
  \item $H_A$: Data \emph{do not} come from the Weibull distributions.
\end{itemize}
The Kolmogorov-Smirnov (KS) test \cite{kolmogorov1933,smirnov1948,corder2009}
was used to decide if simulated data comes from a population with Weibull
distributions, which is based on a comparison between the simulated CDFs in
Fig.\ref{Pileup_PDFCDF_simulation}b and Fig.\ref{Wall_PDFCDF_simulation}b
and theoretical Weibull CDFs from (\ref{eq25}) and (\ref{eq27}).
From Fig.\ref{Weibull_KS} one can see that
p-values are higher than significance level of 0.05 (noted by dash line) usually referred in statistical literature.
The higher values of p-value in the KS-tests mean the higher probabilities of wrong rejection of the fitting $H_0$-hypothesis.
But the p-values higher that the typical significance level ($>0.05$)
do not obligatory mean that the hypothesis is absolutely true.
Rather, the higher p-value for the hypothesis of the Weibull distribution
means the higher probability of wrong rejection of $H_0$-hypothesis
that the estimated distribution follows the Weibull distribution.
It means that $H_0$-hypothesis for the Weibull distribution,
i.e. for pile-ups the simulated data (Fig.\ref{Pileup_PDFCDF_simulation}b) come from the Weibull distribution,
can be accepted in the range of MCSs, where $p_p>0.05$,
and for walls the simulated data (Fig.\ref{Wall_PDFCDF_simulation}b) come from the Weibull distribution,
can be accepted in the range of MCSs, where $p_w>0.05$.
Namely, CDFs of pile-ups (Fig.\ref{Pileup_PDFCDF_simulation}b)
could be Weibull ones for $t>t_{0,p}=10^4$, that agrees with the stated Weibull-like asymptotic (\ref{eq25}) for $t>t_{0,p}$.
And, CDFs of walls (Fig.\ref{Wall_PDFCDF_simulation}b)
could be Weibull ones for $t>t_{0,w}=10^2$, that agrees with the stated Weibull-like asymptotic (\ref{eq27}) for $t>t_{0,w}$.

\begin{figure}[htbp]
\centerline{
    \includegraphics[width=14cm,height=7cm]{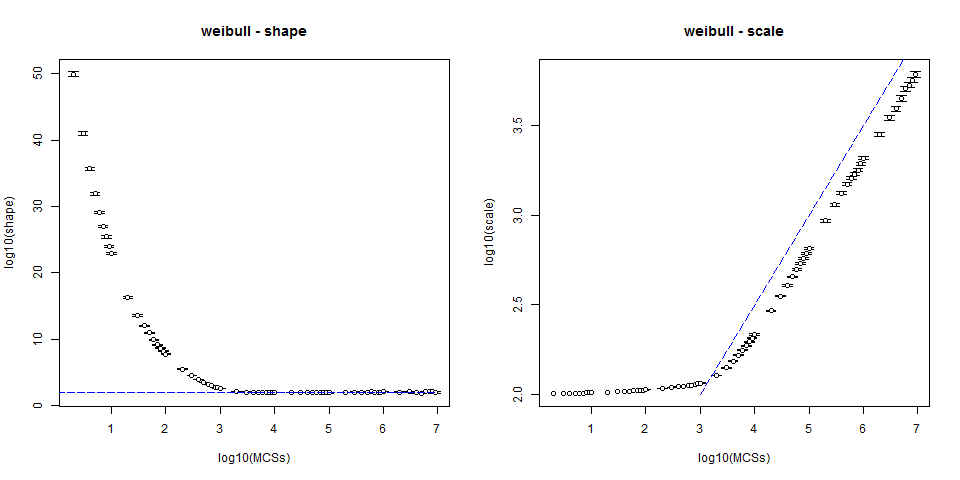}
}
\centerline{a) \hspace{7cm} b)}
\caption{Parameters of Weibull distribution applied for fitting the \textit{pile-up} size distributions: 
(a) shape $ \beta_p$ (the horizontal dashed line denotes 2 on the ordinate axis); 
and (b) scale $\eta_p(t)$ (the slope of the dashed line is 1/2).}
\label{Plieup_Weibull_SHAPE_SCALE}
\end{figure}

According to (\ref{eq24}) fitting pile-up PDFs to Weibull distribution for $t>t_{0,p}$
should give the constant shape parameter, i.e. equal to $\beta_p=2$,
and the following dependency for scale parameter $\eta_p=2(at)^{1/2}$.
This assumption is supported by the pile-up simulated data (Fig.\ref{Plieup_Weibull_SHAPE_SCALE}) for $t>t_{0,p}=10^4$,
where $\beta_p \approx 2$ (the horizontal dashed line in Fig.\ref{Plieup_Weibull_SHAPE_SCALE}a denotes 2 on the ordinate axis)
and $\eta_p \sim \sqrt t$ (the slope of the dashed line in Fig.\ref{Plieup_Weibull_SHAPE_SCALE}b is 1/2).

\begin{figure}[htbp]
\centerline{
    \includegraphics[width=14cm,height=7cm]{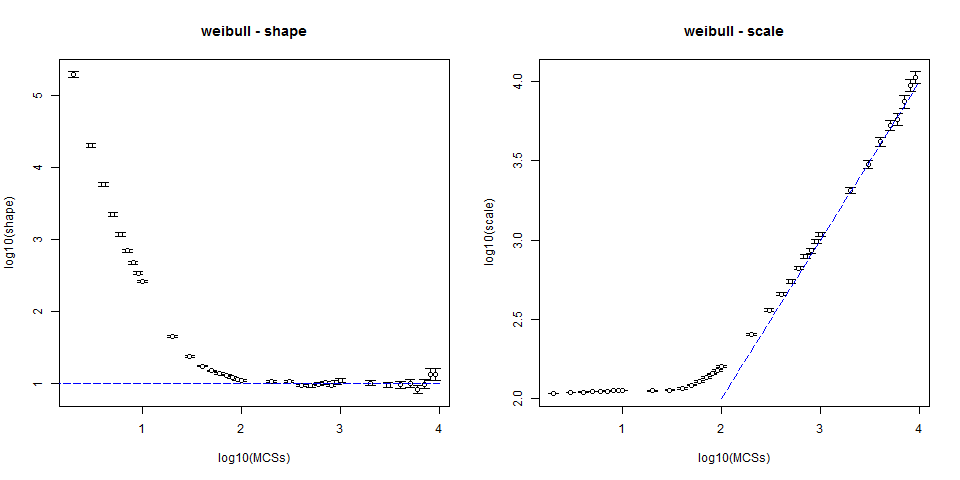}
}
\centerline{a) \hspace{7cm} b)}
\caption{Parameters of Weibull distribution applied for fitting the \textit{wall} size distributions:
(a) shape $\beta_w$ (the horizontal dashed line denotes 1 on the ordinate axis);
and (b) scale $\eta_w(t)$ (the slope of the dashed line is 1).}
\label{Wall_Weibull_SHAPE_SCALE}
\end{figure}

According to (\ref{eq26}) fitting wall PDFs to Weibull distribution for $t>t_{0,w}$
should give the constant shape parameter, i.e. equal to 1,
and the following dependency for scale parameter $\eta_w=at$.
This assumption is supported by the wall simulated data (Fig.\ref{Wall_Weibull_SHAPE_SCALE}) for $t>t_{0,w}=10^2$,
where $\beta_w \approx 1$ (the horizontal dashed line in Fig.\ref{Wall_Weibull_SHAPE_SCALE}a denotes 1 on the ordinate axis)
and $\eta_w \sim t$ (the slope of the dashed line in Fig.\ref{Wall_Weibull_SHAPE_SCALE}b is 1).

In general sense, this means that fitting analysis
(with KS-test and analysis of evolving distribution parameters like shape-scale for the Weibull distributions)
can be used as an additional useful tool
for investigation of any changes of distribution type and related scaling law
(like the boundary-driven change presented here).

\subsection{Moment analysis}\label{section:moment}
The moments like mean $\mu$, variance (square of standard deviation $\sigma$),
excess $\gamma_1$, and kurtosis $\gamma_2$
are of great interest, because they are
measurable in experiments,
are not so vulnerable to fluctuations in empirically constructed PDFs,
follow the general scaling law, and have the clear physical sense.

\begin{figure}[htbp]
\centerline{
    \includegraphics[width=14cm,height=7cm]{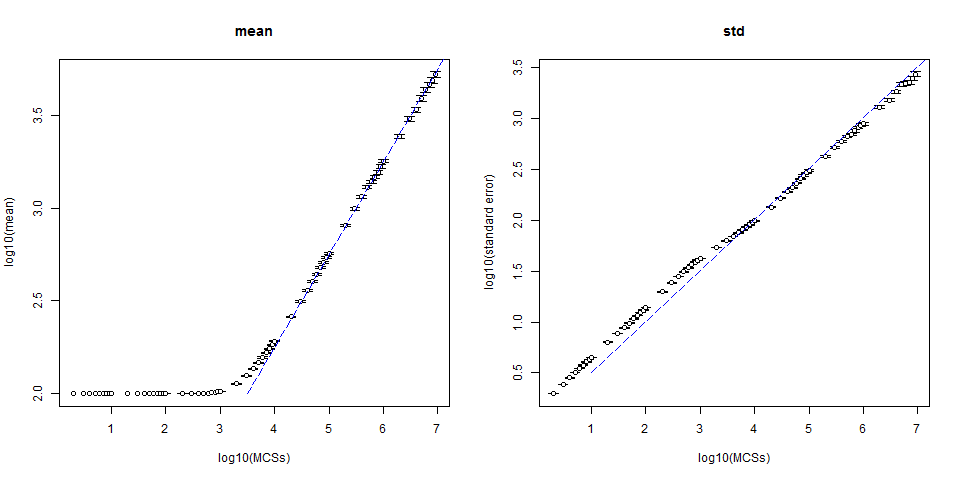}
}
\centerline{a) \hspace{7cm} b)}
\caption{The first two moments for \textit{pile-up} size distributions, represented by:
(a)  mean $\mu_p$ (the slope of the dashed line is 1/2);
and (b) standard deviation $\sigma_p$ (the slope of the dashed line is 1/2).}
\label{Pileup_MEAN_STD}
\end{figure}

\paragraph{Pile-up --- minimum active surface}
For pile-ups,
from (\ref{eq7}) for $t<t_{0,p}$ the mean cluster size is constant $\mu_p=n_0$
and its standard deviation grows like $\sigma_p = 2\sqrt{at}$,
and it is true for the pile-up simulated data (Fig.\ref{Pileup_MEAN_STD}),
where $\mu_p = n_0 = 100$ and $\sigma_p \sim \sqrt t$  for $t>t_{0,p}=10^4$.
But after transition region $t>t_{0,p}$
it follows from (\ref{eq24}) that the mean cluster size is $\mu_p = \eta_p \Gamma[1+1/\beta_p] = \sqrt{\pi at}$
and its standard deviation grows like 
$\sigma_p = \sqrt{\eta_p^2 \Gamma[1+2/\beta_p] - \mu_p^2} = \sqrt{(4-\pi)at}$, 
and it is true for the pile-up simulated data (Fig.\ref{Pileup_MEAN_STD}),
where $\mu_p \sim \sqrt t$ and $\sigma_p \sim \sqrt t$  for $t>t_{0,p}=10^4$.

\begin{figure}[htbp]
\centerline{
    \includegraphics[width=7cm,height=7cm]{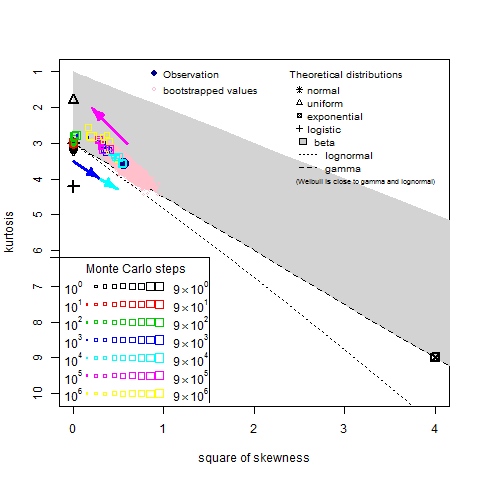}
}
\caption{Moment and bootstrapping analysis of different scaling laws in aggregation kinetics of \textit{pile-ups} presented on the Pearson diagram
in coordinates: kurtosis $\gamma_{2,p}$ and square of skewness $\gamma^2_{1,p}$.
(Color arrows in electronic version are given as eye guides only to follow the movement of squares.)}
\label{Pileup_PearsonDiagram}
\end{figure}

The higher standardized moments of pile-up size distributions, 
namely skewness $\gamma_{1,p}$ and kurtosis $\gamma_{2,p}$,
have the following forms for (\ref{eq24}) for $t>t_{0,p}$:

\begin{eqnarray}
\label{eq28}
    \gamma_{1,p} = (\Gamma[1+3/\beta_p]\eta_p^3-3\mu_p \sigma_p ^2-\mu_p ^3)/\sigma_p ^3, \nonumber 
    \\
    \gamma_{2,p}=(\eta_p^4\Gamma[1+4/\beta_p]-4\gamma_{1,p}\sigma_p^3\mu_p-6\mu_p^2\sigma_p^2-\mu_p^4)/\sigma_p^4
\end{eqnarray}

After simple calculations the square of skewness is equal to
$\gamma^2_{1,p} = 4 (\pi-3)^2 \pi/(4-\pi)^3 \approx 0.4$,
and kurtosis is equal to
$\gamma_{2,p} = (32-3\pi^2)/(\pi-4)^2 \approx 3.25$.
On the Pearson diagram  (Fig.~\ref{Pileup_PearsonDiagram})
\cite{cramer1999mathematical,Rfitdistrplus2012,cullen1999probabilistic}
each data point (square) represents moments (kurtosis $\gamma_{2,p}$ and square of skewness $\gamma^2_{1,p}$)
of the certain simulated PDF for some stage of the aggregation process.
The color of square (in electronic version) denotes exponent,
and the relative size of square (see legend) --- mantissa of MCSs in decimal logarithmic notation.
For example, the smallest red square corresponds to PDF for $t=10$ and the biggest red square --- to PDF for $t=90$.
The data on the Pearson diagram clearly demonstrate the change of the distribution type:
for $t<t_{0,p}=10^4$ more than 30 pile-up PDFs collapse to the initial normal distribution 
(near point with coordinates: kurtosis $\gamma_{2,p}=3$ and square of skewness $\gamma^2_{1,p}=0$)
and after $t>t_{0,p}=10^4$ pile-up PDFs migrate to the zone of Weibull distributions 
(near point with coordinates: kurtosis $\gamma_{2,p} \approx 3.25$ and square of skewness $\gamma^2_{1,p} \approx 0.4$).

\begin{figure}[htbp]
\centerline{
    \includegraphics[width=14cm,height=7cm]{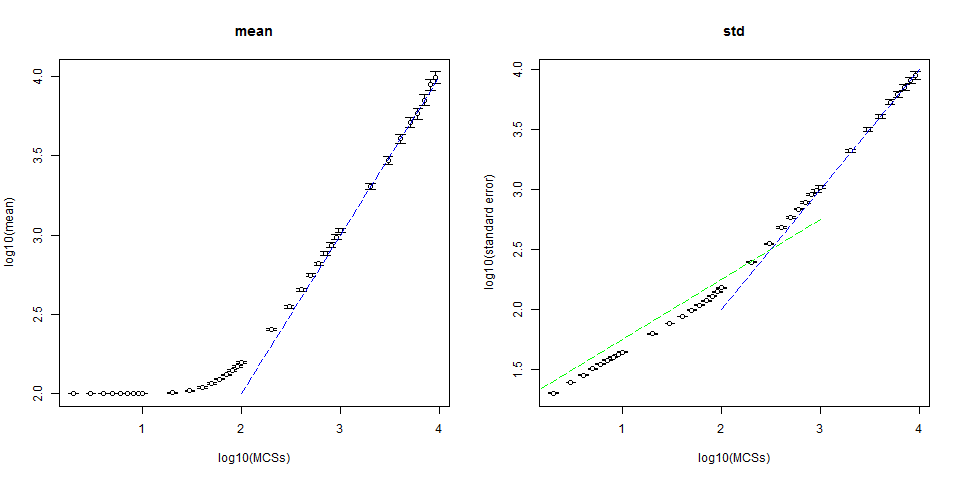}
}
\centerline{a) \hspace{7cm} b)}
\caption{The first two moments for \textit{wall} size distributions, represented by:
(a)  mean $\mu_w$ (the slope of the dashed line is 1);
and (b) standard deviation $\sigma_w$ (the slope of the dashed lines is 1/2 for $t<10^2$, and 1 for $t>10^2$).}
\label{Wall_MEAN_STD}
\end{figure}

\paragraph{Wall --- maximum active surface}
For walls,
from (\ref{eq14}) for $t<t_{0,w}$ the mean cluster size is constant $\mu_w = n_0$
and its standard deviation grows like $\sigma_w \sim \sqrt t$,
and it is true for the wall simulated data (Fig.\ref{Wall_MEAN_STD}) for $t<t_{0,w}=10^2$.
But after transition region $t>t_{0,w}$
from (\ref{eq26}) follows that the mean cluster size $\mu_w = \eta_w \Gamma[1+1/\beta_w] = at$
and its standard deviation should grow like
$\sigma_w = \sqrt{\eta_w^2 \Gamma[1+2/\beta_w] - \mu_w^2} = \sqrt a t$,
and it is true for the wall simulated data (Fig.\ref{Wall_MEAN_STD}) 
where $\mu_w \sim t$ and $\sigma_w \sim t$  for $t>t_{0,w}=10^2$.

\begin{figure}[htbp]
\centerline{
    \includegraphics[width=7cm,height=7cm]{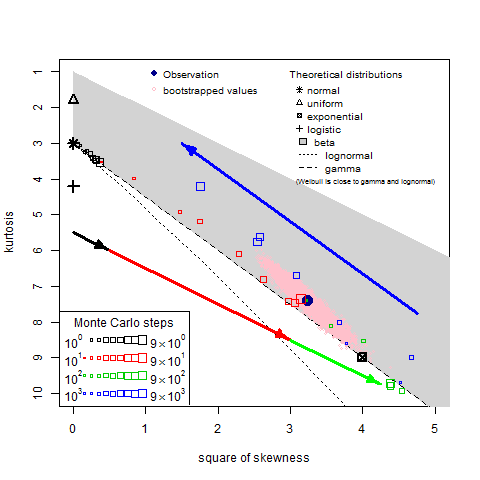}
}
\caption{Moment and boostrapping analysis of different scaling laws in aggregation kinetics of \textit{walls} presented on the Pearson diagram
in coordinates: kurtosis $\gamma_{2,w}$ and square of skewness $\gamma^2_{1,w}$. 
(Color arrows in electronic version are given as eye guides only to follow the movement of squares.)}
\label{Wall_PearsonDiagram}
\end{figure}

The higher standardized moments of wall size distributions,
namely skewness $\gamma_{1,w}$ and kurtosis $\gamma_{2,w}$,
have the similar forms as in (\ref{eq28}).
And after simple calculations the square of skewness is equal to
$\gamma^2_{1,w} = 4$, and kurtosis is equal to $\gamma_{2,w} = 9$.
They are presented on the Pearson diagram  (Fig.~\ref{Wall_PearsonDiagram})
and demonstrate the change of the distribution type: 
for $t<t_{0,w}=10^2$ $<10$ distributions are collapsed to the initial normal distribution
(near point with coordinates: kurtosis $\gamma_{2,w}=3$ and square of skewness $\gamma^2_{1,w}=0$)
and after $t>t_{0,w}=10^2$ they quickly migrate through the zone of Weibull distributions
to the location of exponential distribution denoted by black square with circle and cross inside 
(near point with coordinates: kurtosis $\gamma_{2,w}=9$ and square of skewness $\gamma^2_{1,w}=4$).

The other interesting aspect is the reverse migration of distributions denoted by the backward arrows on the Pearson diagrams:
the magenta arrow for pile-ups in Fig.\ref{Pileup_PearsonDiagram} and 
the blue arrow for walls in Fig.\ref{Wall_PearsonDiagram}.
They correspond to the very late behavior of the simulated systems 
with the limited quantity of particles $10^6$,
when the most part of time the ensemble of large clusters 
exchange by particles without total disappearance of any cluster.
This means clusters do not ``feel'' the 0-boundary and
the third problem with ``infinite'' boundary conditions $-\infty <n<\infty $ 
should be considered for equations (\ref{eq4_heat}) and (\ref{eq10}).
As a result these PDFs become more symmetric (lower skewness) 
and with quasi-normal peaks (lower kurtosis).
But the whole consideration of this third problem is 
beyond the scope of this article and will be considered elsewhere.

In total, the moment analysis allow to identify
the 0-boundary driven changes of distribution type and related scaling law
by the crucial changes of moments, 
like mean $\mu$, variance (square of standard deviation $\sigma$),
excess $\gamma_1$, and kurtosis $\gamma_2$ after transition times $t_0,p$ and $t_0,w$.

\subsection{Bootstrapping analysis}\label{section:bootstrapping}
To check the stability of the simulated distributions shown on the Pearson diagram the bootstrapping analysis
\cite{cullen1999probabilistic}
was applied in order to simulate the exclusion of some clusters from the cluster distribution.
The results of the bootstrapping analysis
for some PDF is plotted on the Pearson diagram
(Fig.~\ref{Pileup_PearsonDiagram} and Fig.~\ref{Wall_PearsonDiagram})
as a pink area of scattered dots.
Each pink dot corresponds to pair of skewness and kurtosis values,
which is calculated for each of many bootstrap samples.

For example, for the pile-up PDF on the Pearson diagram in Fig.~\ref{Pileup_PearsonDiagram}
the results of the bootstrapping analysis for $t=7\cdot10^4$ are plotted as a pink ellipse of scattered dots.
Each pink dot corresponds to pair of skewness and kurtosis values,
which is calculated for each of $>10^4$ bootstrap samples.
It is important to note that
the bootstrapping area for the initial stage $t \ll t_{0,p}=10^4$
has the equiaxial circular shape (it is not shown here) and it is located near the point of the normal distribution:
kurtosis $\gamma_{2,p}=3$ and square of skewness $\gamma^2_{1,p}=0$.
While the bootstrapping area for $t>t_{0,p}=10^4$ is stretched and located in the zone of Weibull distributions
near the point with coordinates: kurtosis $\gamma_{2,p} \approx 3.25$ and square of skewness $\gamma^2_{1,p} \approx 0.4$.

Similarly, for the wall PDF on the Pearson diagram in Fig.~\ref{Wall_PearsonDiagram}
the results of the bootstrapping analysis for $t=10^2$ are plotted as a more elongated pink ellipse of scattered dots.
It is important to note that
the bootstrapping area for the initial stage $t \ll t_{0,w}=10^2$
has the equiaxial circular shape (it is not shown here) and it is located near the point of the normal distribution:
kurtosis $\gamma_{2,w}=3$ and square of skewness $\gamma^2_{1,w}=0$.
And the bootstrapping area for $t>t_{0,w}=10^2$ is stretched
to the location of exponential distribution (black square with circle and cross inside)
near point with coordinates: kurtosis $\gamma_{2,w}=9$ and square of skewness $\gamma^2_{1,w}=4$.

Finally, this means that bootstrapping analysis 
allow to estimate the stability of the estimated moments for the simulated distributions,
determine the areas of their possible locations on the Pearson diagram,
and identify the 0-boundary driven changes of distribution type and related scaling law
by crucial change of the positions and shape of the bootstrapped sample areas in coordinates
$(\gamma^2_{1,p};\gamma_{2,p})$ on the Pearson diagram.

\section{Discussion}

\begin{table}[htbp]
\caption{Comparison of different techniques for identification of the 0-boundary driven change 
of scaling and distribution type at $t\approx t_0$,
where $N_{f_p}$ and $N_{g_p}$ are the numbers of collapsing PDF and CDF curves for pile-ups, 
$N_{f_w}$ and $N_{g_w}$ are the numbers of collapsing PDF and CDF curves for walls,
$p_p$ and $p_w$ are the p-values for KS-test of fitting to Weibull distribution,
$B_p$ and $B_w$ are the locations of the bootstrapped samples in coordinates 
$(\gamma^2_{1,p};\gamma_{2,p})$ on the Pearson diagram.
}
  \label{table:methods}
  \begin{center}
  \begin{tabular}{ccccc}
    \hline
          Example  & Parameter             &             \multicolumn{2}{c}{Value for}                                                                 & No.   \\
                   &                       & $t<t_0$                                          & $t>t_0$                                                &    \\
    \hline
                                                       \multicolumn{5}{c}{\emph{Scaling analysis} (section \ref{section:scaling})} \\
          Pile-ups & $f_p(n,\lambda t)$    & $\lambda^{-1/2} f_{p,t<t_{0,p}}(\lambda^{-1/2} n,t)$ & $\lambda^{-1} f_{p,t>t_{0,p}}(\lambda^{-1/2} n,t)$ & (1) \\
           ---     & $N_{f_p}$             & 18                                               & 30                                                     & (2) \\
           ---     & $g_p(n,\lambda t)$    & $g_{p,t<t_{0,p}}(\lambda^{-1/2} n,t)$                & $g_{p,t>t_{0,p}}(\lambda^{-1/2} n,t)$                                      & (3) \\
           ---     & $N_{g_p}$             & 18                                               & 30                                                     & (4) \\
            \\  
           Walls   & $f_w(n,\lambda t)$    & $\lambda^{-1/2} f_{w,t>t_{0,w}}(\lambda^{-1/2} n,t)$ & $\lambda^{-2} f_{w,t>t_{0,w}}(\lambda^{-1} n,t)$                           & (5) \\
           ---     & $N_{f_w}$             & 9                                                & 18                                                     & (6) \\
           ---     & $g_w(n,\lambda t)$    & $g_{w,t>t_{0,w}}(\lambda^{-1/2} n,t)$                & $g_{w,t<t_{0,w}}(\lambda^{-1} n,t)$                                        & (7) \\
           ---     & $N_{g_w}$             & 9                                                & 18                                                     & (8) \\
            \\       
                                                       \multicolumn{5}{c}{\emph{Fitting analysis} (section \ref{section:fitting})} \\
           Pile-ups & $p_p$ & $<0.05$         & $>0.2$  & (9) \\
           Walls    & $p_w$ & $<0.05$         & $>0.2$  & (10) \\
            \\
                                                       \multicolumn{5}{c}{\emph{Moment analysis} (section \ref{section:moment})} \\
            Pile-ups & $\mu_p$               & $n_0$            & $\sim \sqrt{t}$  & (11) \\
             ---     & $\sigma_p$             & $\sim \sqrt{t}$  & $\sim \sqrt{t}$ & (12) \\
             ---     & $\gamma_{1,p}$    & 0                & 0.632           & (13) \\
             ---     & $\gamma_{2,p}$    & 3                & 3.25            & (14) \\
            Walls    & $\mu_w$               & $n_0$            & $\sim t$         & (15) \\
             ---     & $\sigma_w$             & $\sim \sqrt{t}$  & $\sim t$        & (16) \\
             ---     & $\gamma_{1,w}$    & 0                & 2               & (17) \\
             ---     & $\gamma_{2,w}$    & 3                & 9               & (18) \\
            \\
                                                       \multicolumn{5}{c}{\emph{Bootstrapping analysis} (section \ref{section:bootstrapping})} \\
            Pile-ups & $B_p$ & $\rightarrow$ normal & $\rightarrow$ Weibull   & (19) \\
            Walls    & $B_w$ & $\rightarrow$ normal & $\rightarrow$ exp $\subset$ Weibull   & (20) \\
    
\end{tabular}
\end{center}
\end{table}

Scaling analysis of simulated (or experimentally obtained) size
distributions and histograms (PDFs) is very hard and prone to errors task, because of
the some uncertainty related to proper and optimal binning procedures for
preparation of histograms. In contrast, the scaling analysis of experimental
and simulated CDFs has much higher stability to statistical deviations and outliers among data.
It is partially explained by the absence of shape scaling (\ref{eq16}).
The better collapse of scaled CDF curves could be seen even by visual comparison of
scaled simulated histograms (PDFs) (Fig.~\ref{Pileup_PDFCDF_scaled}a,b and Fig.~\ref{Wall_PDF_scaled})
with their scaled CDFs (Fig.~\ref{Pileup_PDFCDF_scaled}c,d, Fig.~\ref{Pileup_CDF_Gauss_vs_Weibull}, and Fig.~\ref{Wall_CDF_Gauss_vs_Weibull}).

The initial ``singular'' configurations of walls after some time sweeps
evolve to scale-free distributions (Fig.~\ref{Wall_PDF_scaled}),
that is different from pile-up
distributions with broad peaks for the same time (Fig.~\ref{Pileup_PDFCDF_scaled}a).
That is why in practical
sense the notion of ``average cluster size'' seems to be meaningless for
this kind of cluster size distributions (i.e. for exponential-like wall arrangements)
without distinctive peaks. This observations can give some insights as to
possible roots for self-affine arrangements of defects and their manifestations
at the surface of plastically deformed metals.


In general, the boundary driven change of scaling and distribution type at $t\approx t_0$
can be identified by more than 10 parameters summarized in Table~\ref{table:methods}
for 2 examples of aggregation kinetics for clusters with pile-up and wall morphology.
That is why in practice to estimate scale invariant characteristics of systems and their changes
(e.g. the level of hierarchy of defect
substructures and related scale of damage) 
it is necessary to measure and analyze not only the integral ``averaged'' characteristics of defect ensembles
(density per area, average size, etc.), 
but also to define more specific
characteristics, related with their size distributions, like PFDs/CDFs and their parameters \cite{GatsenkoCGW09}.
Moreover, scaling, fitting, moment, and bootstrapping analysis can be very informative and useful 
for the more detailed and precise analysis of evolving distributions.
One of the examples of such scaling analysis applied
for CDFs was demonstrated for statistical analysis of features
observed on surfaces of plastically deformed Al single crystals
under real-time video monitoring and processing \cite{KEM},
Another example of such scaling and statistical analysis 
(Kolmogorov-Smirnov test of fitting, moment analysis, and bootstrapping analysis) 
was proposed for the defect density distribution over the ensemble of nanocrystals. 
And it had shown that change of plastic deformation mode is followed by the qualitative change of defect density distribution type over ensemble of nanocrystals
\cite{Gatsenko2013}.

    It should be noted that the mean cluster size grows
    in agreement with conclusions of nonlinear Leyvraz-Redner scaling theory \cite{Leyvraz}
    about average aggregate $\langle {n} \rangle$ growth like $\langle {n} \rangle \sim t^{1/(2-\alpha)}$.
    Also conclusions for the partial cases (pile-ups and walls) considered here are in agreement with conclusions
    on an average cluster size in Lin-Ke theory for migration-driven aggregation,
    namely pile-ups ($\alpha=0$) grow as $\langle {n} \rangle \sim t^{1/2}$ (diffusive growth \cite{ke2002kinetics})
    and walls ($\alpha=1$) grow as $\langle {n} \rangle \sim t$ (ballistic growth \cite{lin2003kinetics}).
    Also the results stated here are in agreement with scaling estimations of the typical scale growth for diffusive and ballistic regimes
    in Ben-Naim-Krapivsky theory for exchange driven growth \cite{ben2003exchange}
    taking into account the rescaled time variable.

In the more general sense, scaling, fitting, moment, and bootstrapping analysis of experimental distributions
could allow us to determine their intrinsic symmetry
and bring to light the corresponding aggregation scenario for many other
applications related with aggregation phenomena, for example for
investigation of city population dynamics \cite{GatsenkoCGW10}.

At the moment some outliers and unsatisfactory scaling in the simulation data cannot be
compared directly with the theoretical results for large $n$ and $t$ because
simulations were carried out for relatively low number of aggregating
monomers ($\sim $10$^{6})$ and in the limited time range ($<10^7$ MCSs for pile-ups and $<10^4$ MCSs for walls).
That is why to check the exact solutions and make more reliable
conclusions the bigger simulations are under way now in the DG DCI ``SLinCA@Home''
and the science gateway portal on the basis of WS-PGRADE technology\cite{BaskovaIWSG2013}.

\section{Conclusions}
The idealized general model of one-step processes to characterize aggregate
growth was proposed that allowed us to find the scaling characteristics of
some aggregation scenarios. Some factors that can cause scaling transition
and appearance of the self-affine size distribution
of the aggregating system of solitary agents (monomers) and their
aggregates (clusters) were brought to light: the rate of monomer exchange
between clusters, cluster geometry, and initial cluster size. It was shown
that the simplified aggregate growth can be described by the one-variable
Fokker-Planck equation in general form with time-independent drift and
diffusion coefficients. In some aggregation scenarios it could be
transformed to the equivalent equations, from which the non-stationary
analytical solutions with some initial and boundary conditions can be
obtained.

Two primitive cases of cluster aggregation were considered analytically
and simulated by Monte Carlo method to illustrate the different
cluster distributions in different aggregation kinetics: model with
\textit{minimum} active surface (singularity) that corresponds to ``pile-up'' aggregation of
dislocations, and model with \textit{maximum} active surface that corresponds to ``wall''
aggregation of dislocations. It was shown that initial ``singular''
\textit{symmetric} distribution of pile-ups evolves according to ``infinite'' diffusive
scaling law and later it is replaced by the other ``semi-infinite''
diffusive scaling law with \textit{asymmetric} distribution of pile-ups. In contrast, the
initial ``singular'' \textit{symmetric} distributions of walls
initially evolve according to the diffusive scaling law
and later it is replaced by the other linear scaling law with
\textit{scale-free} exponential distributions without distinctive peaks.

    The scaling, fitting, moment, and bootstrapping analysis were proposed for the simulated data
    to identify the 0-boundary driven change of scaling and distribution type at $t\approx t_0$
    by the abrupt change of $>10$ parameters:
    PDF scaling law (No.1,5 in Table~\ref{table:methods}), PDF shape (1,5),
    number and time range of collapsed PDFs (2,6),
    CDF scaling law (3,7), CDF shape (3,7),
    number and time range of collapsed PDFs (4,8),
    p-value in KS-test of fitting CDFs (9,10),
    mean (11,15), standard deviation, which does not change for pile-ups (12,16),
    skewness (13,17), kurtosis (14,18),
    locations of the bootstrapped samples in coordinates on the Pearson diagram (19,20).

From the practical point of view and in the more general sense, 
scaling, fitting, moment, and bootstrapping analyses
of experimental distributions of aggregating monomers and clusters
will allow to determine their intrinsic symmetry, bring to light the
corresponding aggregation scenario, and influence of aforementioned
parameters (rate of monomer exchange between clusters, cluster geometry, and
initial cluster size).

\section{Acknowledgements}
The work presented here was partially funded by the EU FP7 DEGISCO
(Desktop Grids for International Scientific
Collaboration) (http://degisco.eu) project, which is supported by the FP7
Capacities Programme under grant agreement number RI-261561;
EU FP7 SCI-BUS (SCIentific gateway Based User Support) project, contract no. RI-283481;
and partially supported in the framework of the research theme
``Introduction and Use of Grid Technology in Scientific Research of IMP NASU''
under the State Targeted Scientific and Technical Program to Implement Grid Technology in 2009-2013.
The special thanks should be expressed
to the community of volunteer supporter of the the DG DCI ``SLinCA@Home'', 
and especially ``Distributed Computing team ``Ukraine'''' (http://distributed.org.ua)
for their computing resources generously donated for calculation of some of these results.





\bibliographystyle{elsarticle-num}





\end{document}